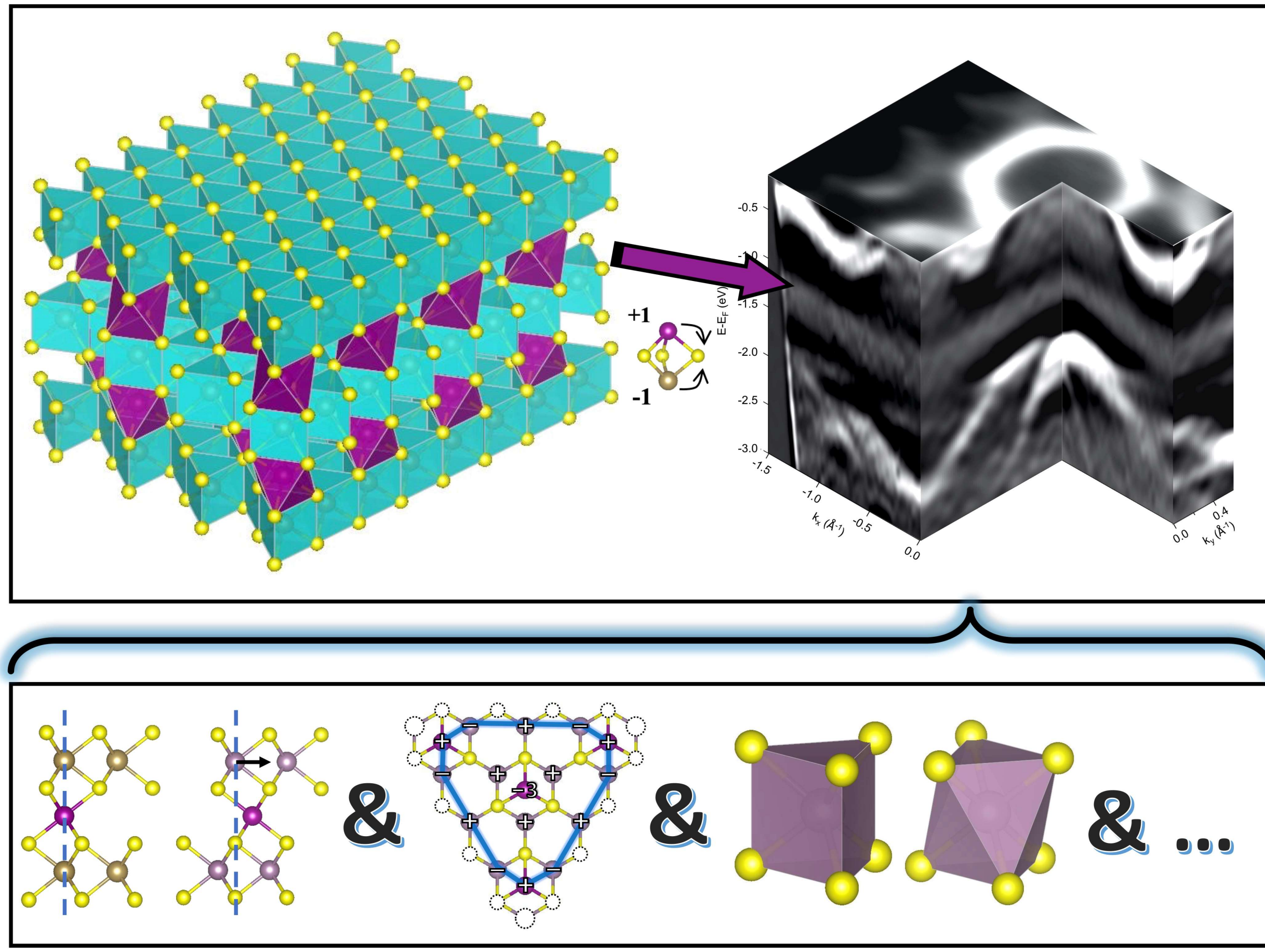

TOC: With a tight binding approach, a theoretical framework is laid down for the universal existence of flat bands in intercalated transition metal dichalcogenides. The transition metal *d*-orbitals form destructive interference via the chalcogenide *p*-orbitals, typically over an extended triangular lattice depending on the nature of intercalation. This concept is experimentally verified in a model system of 2x2 Mn intercalated TaS2.

# Flat Band Generation through Interlayer Geometric Frustration in Intercalated Transition Metal Dichalcogenides

Yawen Peng[1], Ren He[1], Peng Li[1†], Sergey Zhdanovich[2], Matteo Michiardi[2], Sergey Gorovikov[3], Marta Zonno[3], Andrea Damascelli[2], Guo-Xing Miao[1*]

1. Institute for Quantum Computing and Department of Electrical and Computer Engineering, University of Waterloo, Waterloo ON N2L3G1, Canada.
2. Quantum Matter Institute, University of British Columbia, Vancouver BC V6T 1Z4, Canada.
3. Canadian Light Source Inc., 44 Innovation Boulevard, Saskatoon SK S7N 2V3 Canada.

## Abstract

Electronic flat bands can lead to rich many-body quantum phases by quenching the electron's kinetic energy and enhancing many-body correlation. The reduced bandwidth can be realized by either destructive quantum interference in frustrated lattices, or by generating heavy band folding with avoided band crossing in Moiré superlattices. Here we propose a general approach to introduce flat bands into widely studied transition metal dichalcogenide (TMD) materials by dilute intercalation. A flat band with vanishing dispersion is observed by angle-resolved photoemission spectroscopy (ARPES) over the entire momentum space in intercalated $Mn_{1/4}TaS_2$. Polarization dependent ARPES measurements combined with symmetry analysis reveals the orbital characters

* Email address: guo-xing.miao@uwaterloo.ca

† Current address: School of Electronic Science and Engineering, University of Electronic Science and Technology of China, Chengdu, China

of the flat band. Supercell tight-binding simulations suggest that such flat bands arising from destructive interference between Mn and Ta on S through hopping pathways, are ubiquitous in a range of TMD families as well as for different intercalation configurations. Our findings establish a new material platform to manipulate flat band structures and explore their corresponding emergent correlated properties.

## Introduction

Quantum many-body physics with strong electron correlation gives rise to a wide range of exotic electronic properties such as unconventional superconductivity and magnetism[1,2,3,4,5]. This can be achieved in flat band materials characterized by bands with vanishing energy dispersion in momentum space and located near the Femi level. The kinetic energy of electrons is strongly suppressed due to extremely heavy effective mass and gives way to enhanced correlation effects.

The flat band phenomenology has gained significant attention after its experimental realization and observation recently. Except the heavy fermion compounds with 4f electrons, there are generally two types of two-dimensional (2D) materials that host flat bands. One is material with unique frustrated lattice such as the widely studied kagome metal[6,7,8,9,10,11], and the other one is stacked Moiré superlattice such as twisted bilayer graphene (TBG)[14,15,16,17]. The 2D kagome lattice itself has intrinsic topological flat bands due to structural geometric frustration and lattice topology. The ability of tuning flat band to the Fermi level is limited because of its intrinsic property. While the TBG has extrinsic flat band introduced by modifying interlayer interactions and Moiré pattern engineering, with small perturbation to the original Dirac Fermion nature of graphene. By applying electrostatic gating, the Fermi level can be shifted and different filling factor of flat band can be achieved.

Here we report a novel material platform which shows the emergence of flat bands in TMD compounds upon intercalation, different from above two systems. By combining angle-resolved photoemission spectroscopy (ARPES) with density functional theory (DFT) and tight binding calculations, we study the model system of 2×2 $Mn_{1/4}TaS_2$ and establish the ubiquitous existence of flat bands in intercalated TMDs. The origin and property of flat band have similarities with that in kagome metal and Moiré superlattice: while the formation is caused by intrinsic crystal structural factor that the symmetrically aligned Mn and Ta can have destructive wavefunction cancellation on S, the flat band introduced by Mn intercalation has negligible modification to the original band structure of host material $TaS_2$. And the ability of tuning flat band to the Fermi level in such intercalated TMDs can be achieved by varying the intercalant/TMD species and intercalation concentration.

Our ARPES results show a flat band located 1.23eV below the Fermi level, whose orbital characters experimentally determined via polarization-dependent ARPES measurements well agrees with orbital projected DFT calculations as well as crystal symmetry analysis. Furthermore, our tight binding modeling reveals that these flat bands can be generalized to other TMD families and intercalation cases, including H or T phase TMDs, $H_a$ or $H_c$ interlayer stacking sequence, and $\sqrt{3}\times\sqrt{3}$ or any other supercell reconstructions. These findings establish a generic way to introduce and manipulate flat band electronic structures in intercalated TMDs, and shed important light on exploring unique correlated phenomena in these materials.

**Flat bands in $Mn_{1/4}TaS_2$**

The crystal structure of $Mn_{1/4}TaS_2$ (Fig. 1a) contains the host compound 2H $TaS_2$ as well as Mn atoms intercalated into the van der Waals (vdW) gaps, which form an ordered 2×2 periodic sublattice. The successful intercalation is confirmed by X-ray diffraction (XRD, see Supplementary Fig. S1) and high-resolution cross-section transmission electron microscopy (TEM) measurement (Fig. 1b), with intercalated Mn aligned with Ta in the c-direction[12]. Scanning tunneling microscopy (STM) measurements on the (001) cleaved surface show a clear 2×2 periodicity of the Mn layer under the TaS2-terminated surface (Fig. 1c). The measured lattice constant of 2×2 Mn (Supplementary Fig. S1) is 6.6 Å which is twice that of $TaS_2$.

To study the electronic structure of $Mn_{1/4}TaS_2$, we start by performing the DFT calculation and tight binding simulation. The DFT calculated band structure without spin polarization mainly consists of host $TaS_2$ dispersion with additional folded bands due to the lager supercell. Several flat bands are observed near the Fermi level which are mainly from Mn 3d orbitals (Supplementary Fig. S1). To understand the origin of the flat band, a 9-band tight binding Hamiltonian with only s-orbitals and nearest neighbor hopping is constructed (see Supplementary Note SI). The toy model contains 4 Ta, 4 S and 1 Mn atoms, forming a honeycomb-like Ta-S layer with 2×2 Mn stacked on top of Ta and connected via S (Supplementary Fig. S2), which is the building block of the unit cell. This s-orbital tight binding model can be generalized to d-orbital one, because each Mn has a Ta symmetrically aligned with respect to the S plane, therefore sharing same hopping phases regardless s or d orbitals.

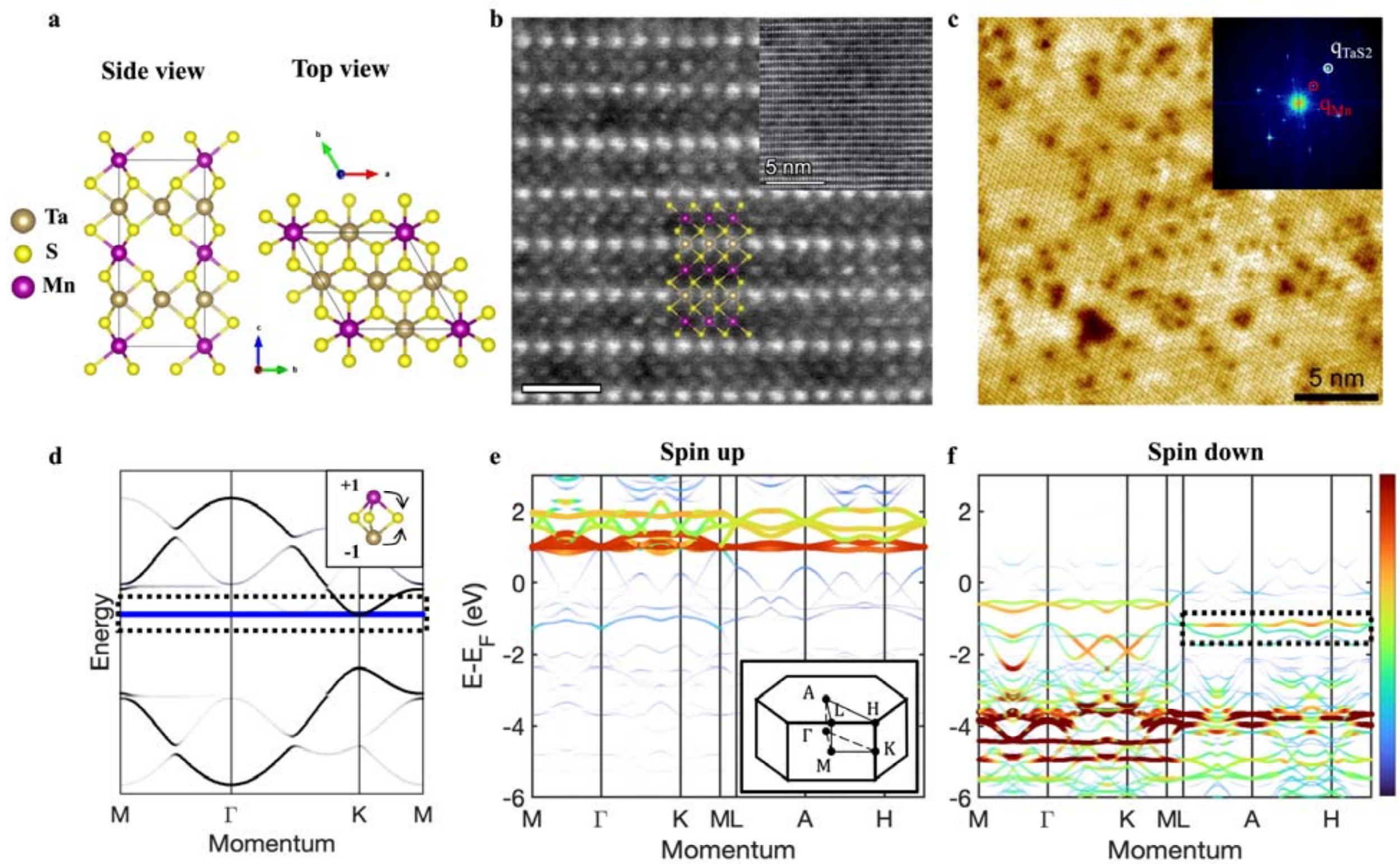

Fig. 1: Crystal characterization and band structure calculation. **a**, Schematic of crystal structure of $Mn_{1/4}TaS_2$. **b**, STEM cross-section view of $Mn_{1/4}TaS_2$ along [100] axis. The crystal structure is overlaid with the image. Scale bar is 1nm. The inset is the large-scale TEM image with scale bar 5 nm. **c**, STM atomic resolution of $TaS_2$ termination ($V_{bias}$ = 1 V, $I_{set}$ = 0.5 A). Scale bar is 5nm. The inset is the corresponding FFT pattern. White and red circles denote 1×1 $TaS_2$ and 2×2 Mn lattice. **d**, Unfolded band structure of 9-band tight binding model (same parameters used in Supplementary Fig. S2). The flat band with blue color is indicated by dashed box. The inset shows the localization due to destructive interference on S. The amplitudes are not necessary to be equal, depending on Mn-S/Ta-S hopping

difference. **e**,**f**, DFT band structures of $Mn_{1/4}TaS_2$ with **e** spin up and **f** spin down components. The line width and color represent flat band weight projections of Mn. Inset in **e** is the 3D Brillion zone. The observed flat band in ARPES is indicated by the dashed box in **f**.

The tight binding band structure is given in Fig. 1d. Two dispersive bands (black-grey curves) come from the original honeycomb-like Ta-S lattice with gap opening at K point, while the flat band (blue curve) originates from the destructive interference between Mn and Ta orbitals (Fig. 1d inset). The dilute intercalation ensures no interaction between Mn-Mn lattices (set to zero in this toy model), while the hopping between Mn-S is destructively cancelled by that of Ta-S directly below. This makes wave function localized in the Mn-S-Ta trigonal bipyramidal structure and cannot propagate beyond S edges.

The flat band is strongly related to the onsite potential of Mn ($\varepsilon_{Mn}$) and will move accordingly when $\varepsilon_{Mn}$ changes (see Supplementary Note SI and Fig. S3). The simplified tight binding model is consistent with the spin-polarized DFT calculation (Fig. 1e,f). The flat bands mostly have contributions from Mn d-orbitals. Due to the exchange splitting of Mn 3d electrons, the flat bands split into two groups with opposite spin polarizations. The spin up component has higher energy which is closer to the onsite potential of the Ta 5d electrons according to Wannier calculations, so this set of flat bands moves upward above the Fermi level and becomes flatter. On the contrary, the spin down component has lower energy and larger energy difference with the Ta 5d orbitals, so these flat bands move downward and become more dispersive, affected by hybridization with the S bands as well.

The existence and the origin of flat bands in $Mn_{1/4}TaS_2$ revealed via DFT and tight-binding simulations are further confirmed by ARPES measurements presented in Fig. 2. Due to the supercell folding, the pristine Brillouin zone is reduced into a smaller hexagon as shown in Fig. 2a. The two hexagonal barrels centered at the Gamma point and the two ring-shaped barrels centered at the K points are the typical features of the host 2H $TaS_2$ with spin-orbital coupling (SOC) splitting. However, looking closely, there exist additional weaker arcs caused by the folding of these original $TaS_2$ bands. The barrels around the K points are folded to around the Gamma point and form a new hexagonal electron pocket. In addition, the Gamma point barrels are folded to M points, forming several arcs that connect barrels at different K points on the Fermi surface.

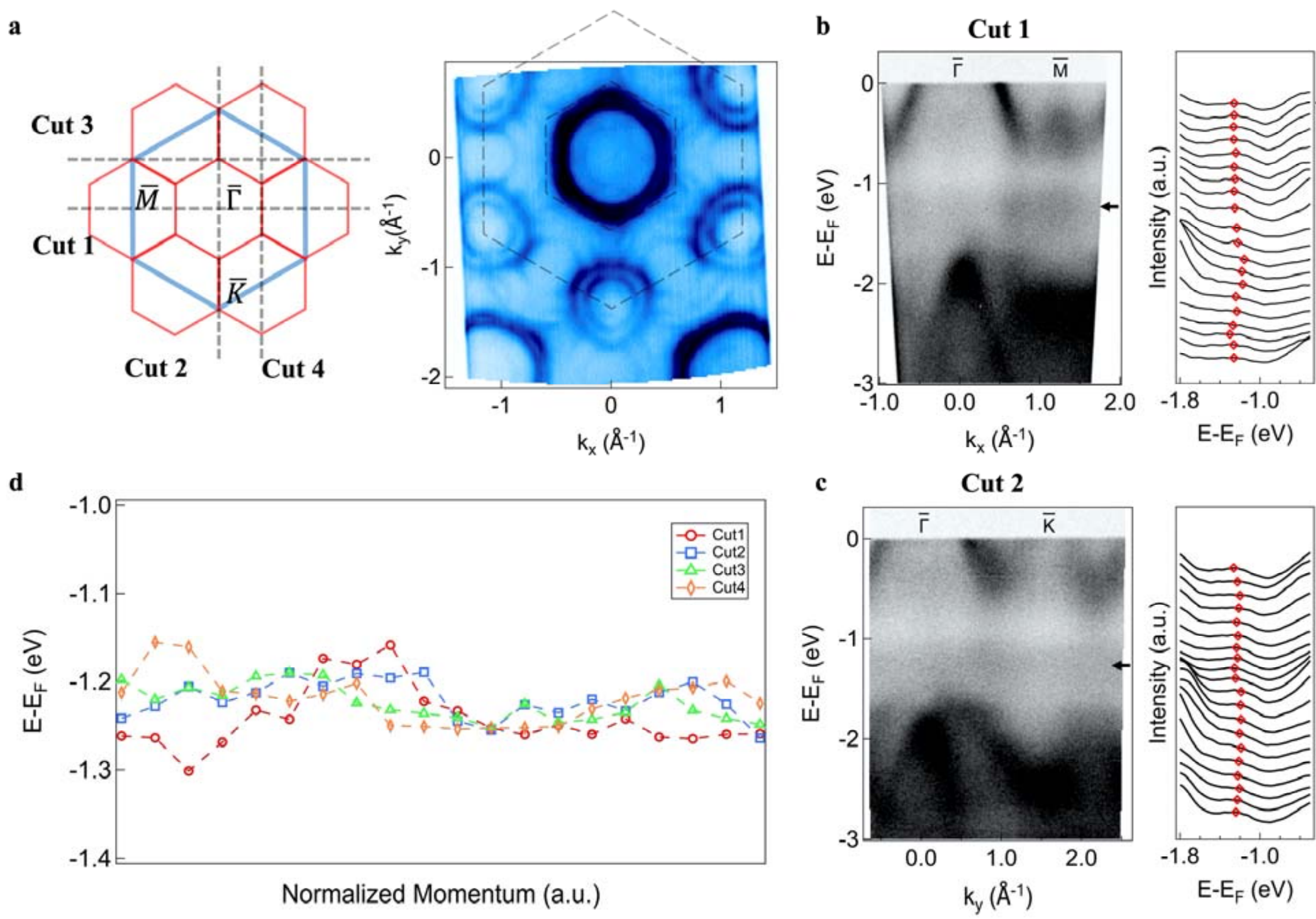

Fig. 2: Flat bands in $Mn_{1/4}TaS_2$ revealed by ARPES. **a**, (Left) Schematic of two-dimensional Brillouin zone of $Mn_{1/4}TaS_2$ with high-symmetry points labeled. The blue hexagon represents the primitive $TaS_2$ Brillouin zone and blue hexagons represent the reduced ones of intercalated $Mn_{1/4}TaS_2$. Dashed lines indicate ARPES momentum-space cuts directions in **b** and **c**, as well as in Supplementary Fig. S4. (Right) Fermi surface of $Mn_{1/4}TaS_2$. Dashed lines indicate the primitive and reduced Brillouin zones. **b**,**c**, Different ARPES spectra (Cut 1 and 2) with corresponding EDCs along high-symmetry directions. Black arrows mark the flat band energy positions, and red rhombus dots track the flat band peaks in EDCs. **d**, Evolutions of the peak positions in Cut 1-4. The black solid line indicates the approximate position of the flat band. All the data were acquired with linear horizontal photons with an energy of 75 eV.

Different cuts across the Brillouin zone are examined and a flat band is identified over the whole momentum space. Fig. 2b and 2c give the ARPES spectra measured along high-symmetry paths $\bar{\Gamma}-\bar{\mathrm{M}}$ and $\bar{\Gamma}-\bar{\mathrm{K}}$ respectively. The flat bands indicated by black arrows fall right inside the $TaS_2$ gap formed between bands crossing the Fermi level with primarily Ta $d_{z2}$ nature (-0.7eV and above) and those with primarily S p orbitals natures (-1.5eV and below). To better visualize the dispersionless band structures, the stackings of energy distribution curves (EDCs) for different cuts are presented (Fig. 2b,c). The flat bands manifest as peaks whose positions are tracked by Lorentzian fittings (see ARPES EDC peak fitting in Experimental Section and Supplementary Table. S1) of the integrated EDCs. Other spectra taken across high-symmetry points (Cut 3 and 4 along the directions shown in Fig 2a) are examined in Supplementary Fig. S4. The second derivative plots of all these cuts are provided as well (Supplementary Fig. S5), where the flat bands have improved visibility and get better resolved. The evolutions of flat band peak positions from the fittings are summarized in Fig. 2d. The flat bands are located around 1.23 eV below the Fermi level, and they exhibit a negligible dispersion in energy throughout the whole momentum space, again confirming the flat band nature in the $Mn_{1/4}TaS_2$. The bandwidth of the flat band is estimated to be 0.15 eV. Broadening can be attributed to the presence of NNN hopping (see Supplementary Note SII and Fig. S3) and a $k_z$ dispersion due to interlayer hopping (see Supplementary Note SV, Fig. S12 and Fig. S13). Overall, the ARPES measurements provide experimental evidence of flat bands in this intercalated TMD system, and consistent with both the theoretical DFT calculation and the simplified tight binding modeling.

**Polarization dependent ARPES measurement**

In addition to the existence of flat bands, we noticed that there are some variations of their spectral intensities across momentum space. To better understand the origins and properties of the flat bands, the ARPES band dispersion measurements along the $\bar{\mathrm{M}}-\bar{\Gamma}-\bar{\mathrm{M}}$ and $\bar{\mathrm{K}}'-\bar{\mathrm{K}}-\bar{\mathrm{K}}'$ (which is parallel to $\bar{\mathrm{M}}-\bar{\Gamma}-\bar{\mathrm{M}}$) directions are performed with both linear vertical and horizontal polarizations of the incident light. In the linear horizontal (LH) polarization, the flat band is suppressed in intensity near Gamma point (Fig. 3a). As a comparison, when the incident photon is linear vertical (LV) polarized, the spectral intensity near Gamma point becomes enhanced. This feature gets resolved more clearly in the momentum distribution curve (MDC) at the flat band positions. Fig. 3b gives the comparison of MDCs extracted from both ARPES spectra: the dip near Gamma point observed with LH polarization (red curve) transforms into a hump in LV polarization (blue). The intensity asymmetry in LH polarization is due to the matrix element effect, that the photoemission intensity depends on the angle between the momentum of electron and the electric field vector of photon. Same behavior of spectra intensity variation is found at K point as well (Supplementary Fig. S6). The overall flat band intensities compensate for each other and fill up the whole momentum space (see Supplementary Fig. S7).

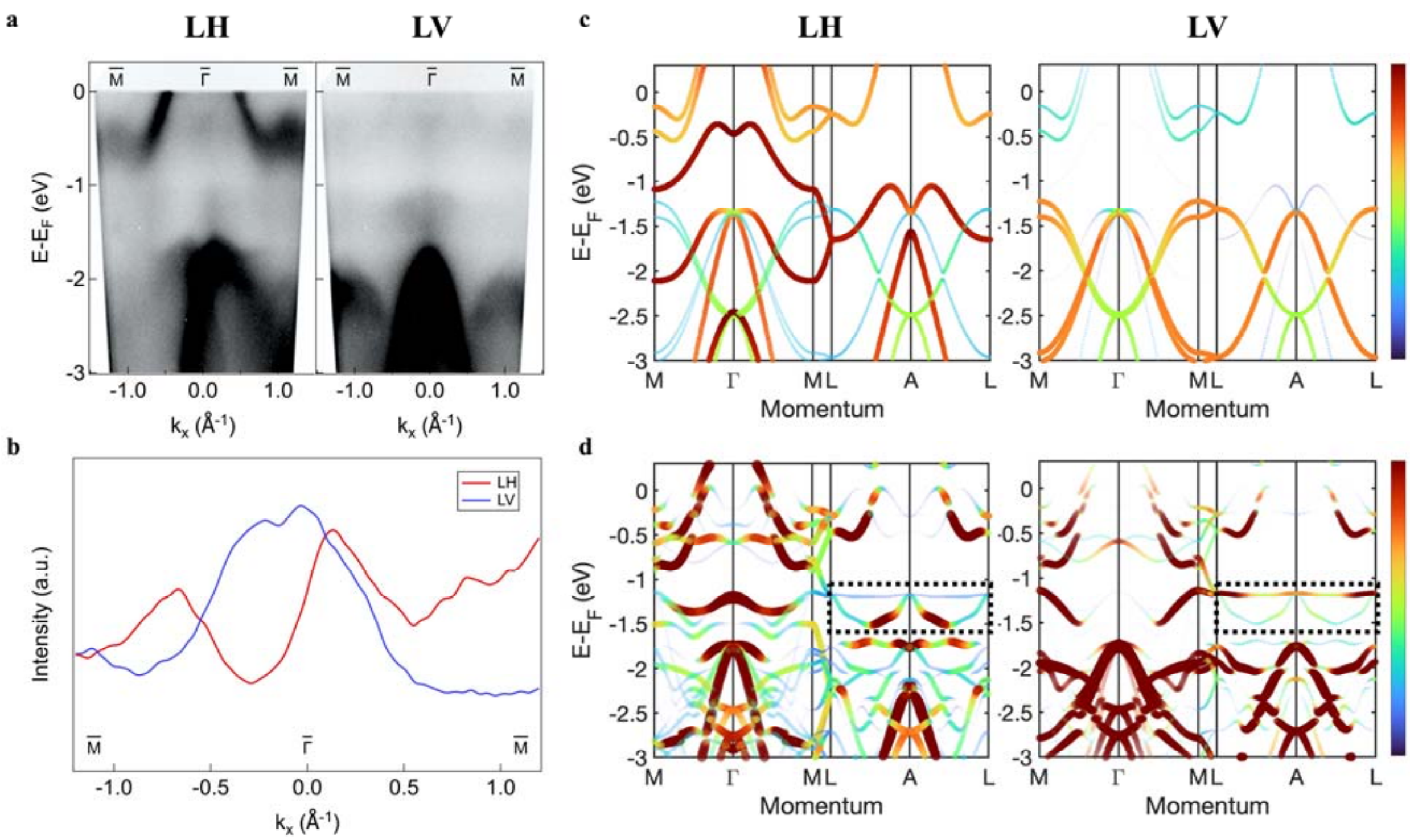

Fig. 3: Polarization dependence and orbital characters of flat bands. **a**, ARPES spectrums measured along $\bar{\mathrm{M}} - \bar{\Gamma} - \bar{\mathrm{M}}$ direction with (left) LH and (right) LV polarization. **b**, MDCs at the flat band position (integrated over an energy range of 40 meV around -1.23 eV) in (red) LH and (blue) LV polarizations. **c**, DFT band structures of $TaS_2$ along M-Γ-M/L-A-L directions with LH and LV polarizations respectively. Line width and color represent even/odd orbital weight projections in LH/LV polarization. **d**, Projected DFT band structures of 2×2 $Mn_{1/4}TaS_2$ with band unfolding in LH/LV polarizations. Dashed boxes indicate the flat band positions.

This intensity variation is caused by the dipole selection rules in linearly polarized photoemission and can be used to reveal the orbital characters of the flat bands. In the experimental set-up used for our polarization dependent ARPES measurements (Supplementary Fig. S8), the LH (LV) polarization has even (odd) parity with respect to the mirror plane (the beam incident xz plane in the illustration), which means only those bands that have the same even (odd) parity can be selectively resolved. Then the d and p orbitals can be divided accordingly to their parity into two groups which are $d_{z2}$, $d_{x2-y2}$, $d_{xz}$, $p_z$, $p_x$ and $d_{xy}$, $d_{yz}$, $p_y$ respectively (Supplementary Fig. S8). Orbital projected calculations along the M-Γ-M/L-A-L directions are further performed to compare with the experimental results. We first examine the $TaS_2$ band structure and compute the orbital projections under LH/LV polarizations (Fig. 3c) to show good agreement with experiments. With LH polarization, the dominant features coming from host $TaS_2$ are hole pocket (Ta $d_{z2}$) above aforementioned gap and X-shaped bands (S $p_z/p_x$) below. They get strongly suppressed and instead two electron and hole pockets (S $p_y$) are observed in LV polarization.

Based on the selection rules and orbital projections, we further check the parities of flat band in $Mn_{1/4}TaS_2$. It is worth noting that there are three sets of flat bands according to the calculation, two of which are located at relatively constant energy while one has noticeable dispersion (Fig. 1e and Supplementary Fig. S1). According to the crystal field analysis, Ta in the H phase $TaS_2$ is in a trigonal prismatic coordination with local $D_{3h}$ symmetry, while the Mn intercalants occupy the octahedral-like interstitial sites and have local $D_{3d}$ symmetry, the same as T phase TMD. Both $D_{3h}$ and $D_{3d}$ symmetries have three groups of irreducible representations, thus five degenerate d orbitals split into $d_{z2}$, $d_{x2-y2}/d_{xy}$ and $d_{xz}/d_{yz}$ respectively due to crystal field splitting[21,22]. Among these, the $d_{z2}$ orbitals are more susceptible to NNN hopping in the z-direction therefore are relatively more dispersed. In tight binding simulation (see Supplementary Note SI), the flat band position is strongly affected by onsite potentials, therefore the observation of three flat bands in orbital projected calculation is consistent with the symmetry analysis and tight binding modeling.

The $d_{x2-y2}/d_{xy}$ ($d_{xz}/d_{yz}$) flat bands are doubly-degenerate and the two orbitals have opposite parities with respect to the mirror plane, thus polarization-dependent spectral intensity variation in the ARPES measurements is expected. The overall orbital projected DFT calculation of $Mn_{1/4}TaS_2$ (Fig. 3d) is further examined. Contributions to the flat band observed in ARPES mainly come from Mn $d_{xz}/d_{yz}$ and Ta $d_{z2}/d_{x2-y2}/d_{xy}$ orbitals, and S $p_z/p_x/p_y$ spectral weights become more pronounced for flat bands of spin down Mn, due to their lower onsite potentials and stronger hybridization. The calculation shows the same weight distribution of flat bands, where the projected orbitals have stronger spectral intensity at Gamma point under LV polarization and weaker intensity under LH polarization. The result is in good agreement with ARPES measurements and reveals the orbital characters of flat bands in $Mn_{1/4}TaS_2$.

**Flat bands in other intercalated TMDs**

The flat band in $Mn_{1/4}TaS_2$ is identified experimentally by ARPES measurements and theoretically by DFT calculation as well as tight binding modeling. From here, the ubiquitous existence of flat bands in a wider collection of intercalated TMD is examined, and the phenomenon is shown to be generic and generalizable to other TMD families.

In a TMD with $2H_a$ structure like $Mn_{1/4}TaS_2$, the transition metal (TM) atoms are aligned in the c direction. From TEM results and first principle relaxations, the intercalants in the vdW gap are aligned with the TM atoms in adjacent layers (Fig. 4a left panel, stacking ABA”b”CBC…). However, the intercalation can be quite different in $2H_c$-TMD materials like $MoS_2$[31]. In the unit cell of a $2H_c$ structure, there is an in-plane sliding between layers, and the TM atoms of one layer are no longer aligned with those of the next layer, instead aligned with the chalcogen atoms. The intercalants between layers now sit in the hollow sites of the honeycomb-like TMD lattice (Fig. 4a right panel, stacking ABA"c"BAB…). This is more like a buckled and dilute version of Dice lattice, which is a well-known frustration lattice that can host flat bands[18,19,20]. The Dice lattice realization can be achieved in intercalated $2H_c$-TMD, because intercalants have strong hopping with chalcogen atoms but negligible hopping with TM atoms due to the extension into 3D layered structure.

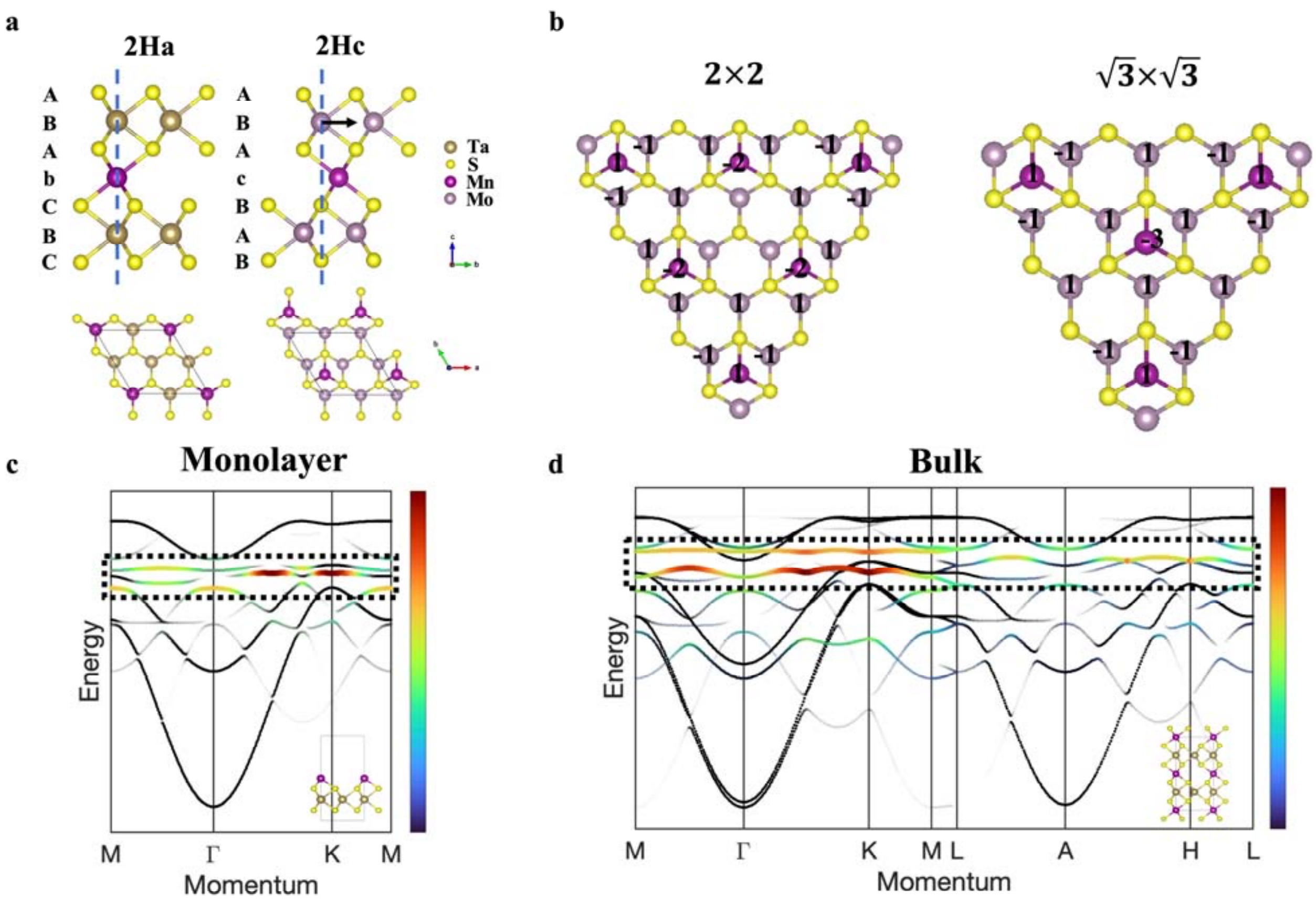

Fig. 4: Flat bands in other sparsely intercalated TMD. **a**, Side and top view of intercalation structure of $2H_a$- and $2H_c$-

TMD. The dashed lines mark the interlayer alignment of TM atoms with respect to chalcogen atoms. “A”, “B/b” and “C/c” indicate the layer stacking sequences and intercalant positions. **b**, Localized states in $2H_c$ structures of (left) 2×2 and (right) $\sqrt{3} \times \sqrt{3}$ with amplitude/phase of each atom labeled. To simplifiy the hopping parameters are set the same. **c**,**d**, Tight binding band structures in **c** monolayer and **d** bulk intercalated $2H_a$ TMD (same parameters used in Supplementary Fig. S14). Colors represent the Mn flat bands (indicated by dashed boxes) weight projections. Insets are tight binding model structures.

Although the 2×2 supercell tight binding simulation shows the same band dispersion as well as flat band existence in intercalated $2H_a$ and $2H_c$ structures, the underlying localization mechanisms are quite different[13]. The localized state in the $2H_a$ structure is confined in the trigonal bipyramid structure (Fig. 1d), and the wavefunction cancellation happens between the intercalant and its neighboring aligned TM atom through chalcogen atoms. But this becomes more complicated in the $2H_c$ structure (Fig. 4b). The environment of an intercalant involves three neighboring TM atoms, and the localized state is geometrically confined in a large triangle that encloses 6 intercalated atoms. The amplitudes and phases of each atom are indicated and they form destructive interference on the chalcogen atoms along the perimeter. This prevents any wavefunction propagation out of the triangle, leading to the electronic localization and flat bands with quenched kinetic energy.

More situations for intercalation concentrations and supercell structures are examined. It is not hard to find that the flat band exists in $2H_a$ structures, independent on the supercell geometry even for random dilute intercalations. This is because the localization only involves the intercalant and its neighboring aligned TM atoms. The situations are quite different and complex for the $2H_c$ structures. Although the flat band still exists in these structures, the geometry of localization is supercell dependent. The localized states of $\sqrt{3}\times\sqrt{3}$ supercell is presented in Fig. 4b and other commonly occurring cells are summarized in Supplementary Fig. S9. The 1×1 primitive cell is indeed a Dice lattice and other supercells can be viewed as generalized cases. For a $n \times n$ supercell, the geometry of the localized state always consists of a $(n^2+1)\times(n^2+1)$ triangle without the very corner TM atoms. Clearly the localization is more extended in real space with lower intercalation concentration. The triangle orientation is aligned with the host TMD lattice, and the corners always start and end with intercalants. For an arbitrary intercalation ratio forming an ordered supercell in $2H_c$-TMD, we can always construct such localized triangular electron pockets and achieve flat bands. When there is more than one type of intercalant in one unit cell, or supercells with different sizes are mixed, the localization can also be realized by the superposition of the individual localized pockets (Supplementary Fig. S10), and flat bands are still present.

We further extend the tight binding simulation to monolayer and bulk intercalated $2H_a$-TMD models respectively. The monolayer structure (Fig. 4c) contains a single TMD layer with 2×2 intercalants on the surface in the slab model, while the bulk structure (Fig. 4d) has two layers in $2H_a$ stacking with intercalants aligned with TM atoms in the unit cell. The tight binding calculation (with NNN hopping of Ta-Ta, S-S and Ta-Mn taken into consideration as well as $k_z$ dispersion) again reveals the existence of nearly flat bands in both structures. The observed deviation from flatness comes from the not perfect cancellation on the S sites in the monolayer case, and from the $k_z$-dependent interlayer hopping along the c direction in the bulk case (see Supplementary Note SV and Fig. S11, 12). The T phase TMD and other ways of intercalant stacking are further explored (Supplementary Fig. S14). The nearly dispersionless features are preserved among all categories discussed above, revealing the wide generality of flat band existence in the intercalated TMD family.

**Conclusion**

In conclusion, we provide experimental and theoretical evidence for the existence of flat bands in the model intercalated TMD system $Mn_{1/4}TaS_2$. The weak direct interaction between adjacent Mn atoms, and the destructive interference between Mn and Ta on S contribute to the electron wavefunction localization and vanishing energy dispersion. With tight binding modeling, we further provide a generic way to introduce flat bands into TMD family by intercalation. The position of flat bands on the energy scale can be tuned by varying intercalant species and ratios, to achieve different onsite potentials and crystal field/spin exchange splitting shifts. Moreover, strong spin-orbital coupling interaction can be introduced from either host TMD compounds or intercalants with d electrons, to explore potential topological features[30]. Therefore, the criteria for flat bands in such systems comprise (1) dilute intercalation to eliminate direct hopping between intercalants and reduce dispersions from influences like NNN hopping; (2) successful intercalation rather than random substitution or interstitial defects to maintain the destructive interference, and

(3) closer onsite potentials of the intercalants and the host transition elements for more ideal electron wavefunction cancellation. In the meantime, the choice of onsite potentials is also a powerful way to adjust the flat band position with respect to the Fermi level. The construction of flat bands provides a unique route for engineering electron correlations in TMDs, and for the systematic exploration and manipulation of the rich properties within this class of materials.

**Acknowledgement**

This work was supported by the Natural Sciences and Engineering Research Council of Canada (NSERC) Discovery Grant RGPIN-03753 and the Canada First Research Excellence Fund – Transformative Quantum Technologies. The work at UBC was undertaken thanks in part to funding from the Max Planck-UBC-UTokyo Centre for Quantum Materials and the Canada First Research Excellence Fund, Quantum Materials and Future Technologies Program. This project is also funded by the Canada Foundation for Innovation (CFI); the British Columbia Knowledge Development Fund (BCKDF); the Department of National Defence (DND); and the CIFAR Quantum Materials Program (A.D.). Use of the Canadian Light Source (Quantum Materials Spectroscopy Centre), a national research facility of the University of Saskatchewan, is supported by CFI, NSERC, the National Research Council, the Canadian Institutes of Health Research, the Government of Saskatchewan, and the University of Saskatchewan. Support by CLS Rapid-Access Funding is also acknowledged.

**Experimental Section**

Single crystal growth

Single crystals of $Mn_{1/4}TaS_2$ were grown using the chemical vapor transport (CVT) technique, with iodine added as the transport agent. A mixture of Mn (99.9%), Ta (99.9%) and S (99.9%) with a nominal stoichiometry of $Mn_{1/4}TaS_2$ was ground and placed into a quartz tube, followed by the introduction of iodine into the mixture. The quartz tube was then evacuated to remove any air or gas. Subsequently, the tube was subjected to a thermal treatment in a gradient from 1260 K to 1170 K for 10 days. Upon opening the tube, the obtained crystals were cleaned using supersaturated aqueous solutions of KI for ultrasonication, followed by washing with deionized water and alcohol. The crystals have dimensions of up to several mm in diameter and exhibit a hexagonal morphology.

Sample preparation and characterization

STM measurements were carried out in a commercial Omicron LT-STM with ultrahigh vacuum (base pressure better than $1 \times 10^{-10}$ mbar). Single crystal flakes with regular hexagonal shapes were mounted on the STM sample holders with conducting epoxy H21D and heated up to 150 degree for 15 minutes. Then ceramic posts were attached on the flakes' surfaces with epoxy H21D and the samples were heated at 150 degree for another 45 minutes. The samples were loaded into STM and cleaved *in-situ* by a wobble stick to expose fresh surfaces for the measurements at 77K (the cleaving method in ARPES is the same as in STM). A tungsten tip was used and calibrated on Au (111) before the measurement. Single-crystal XRD patterns were collected with a D8-VENTURE-XRD diffractometer with a 2-bounce Ge (022) monochromator and Cu Kα line ($\lambda$ = 0.15418 nm). STEM samples were prepared by focus ion beam (FIB) with a Zeiss Auriga 40 SEM/FIB and cross-section high-angle annular dark-field (HAADF) images were collected with a TFS Spectra Ultra.

## ARPES measurements

The ARPES measurements were performed at the Quantum Matter Spectroscopy Center (QMSC) beamline at the Canadian Light Source. The samples were cleaved *in-situ* at pressures better than $10^{-11}$ Torr. The measurements were performed at base temperature of 15 K using a Scienta R4000 hemispherical analyzer equipped with a horizontal entrance slit. Overall angle and energy resolution were better than 0.1° and 23 meV respectively. Photon energy 75 eV was used with varying linear horizontal and linear vertical polarizations.

## DFT and Wannier calculations

All first-principles DFT calculations were implemented in the QUANTUM ESPRESSO package[25,26] using the Perdew-Burke-Ernzerhof exchange-correlation functional[27]. A cut-off energy of 780 eV was used for the plane-wave basis set and a k-point mesh of 6×6×4 was applied for the 3D Brillouin zone sampling. The structures were fully relaxed until the residual force on each atom was under 0.01 eV Å-1. Hubbard $U = 2$ eV on Mn d orbitals was used for band structure calculations. The Wannier calculations including 98 orbitals (Mn: 3d, Ta: 5d and S: 2p) were fitted from DFT results with Wannier functions using the Wannier90 package code[28,29].

Supercell band unfolding was performed by calculating the spectral weights at each k-point in the primitive Brillouin zone[23,24].

ARPES EDC peak fitting

ARPES intensity can be expressed in terms of the product of interaction matrix element $M$, the Fermi-Dirac distribution function $f$ and the single-particle spectral function $A$:

$$I(\boldsymbol{k},\omega) = M(\boldsymbol{k},\omega)f(\omega)A(\boldsymbol{k},\omega)$$

The matrix element term $M$ can be approximated as a constant and the Fermi-Dirac distribution function $f$ describes almost occupied states below the Fermi level at low temperature. Therefore the intensity is proportional to the single-particle spectral function $A$ which has a Lorentzian function form:

$$A(\boldsymbol{k},\omega) = -\frac{1}{\pi}\frac{\Sigma''(\boldsymbol{k},\omega)}{[\omega - \epsilon_0(\boldsymbol{k}) - \Sigma'(\boldsymbol{k},\omega)]^2 + [\Sigma''(\boldsymbol{k},\omega)]^2}$$

To track the peaks of flat bands in EDCs with given momentum, a Lorentzian fitting with background is applied:

$$I(E) = \frac{A}{\pi}\frac{\Gamma/2}{(E - E_0)^2 + (\Gamma/2)^2} + B(E)$$

where $A$ is the amplitude of the spectral function, $E_0$ is the peak position, $\Gamma$ is the full width at half maximum (FWHM) of the spectral function and $B(E)$ is the cubic polynomial background $B(E) = b_0 + b_1E + b_2E^2 + b_3E^3$.

**Supplementary Notes:**

## SI. Tight binding model for intercalated TMD

The simplified tight binding structure uses an artificial monolayer which is the building block of $Mn_{1/4}TaS_2$ unit cell, containing only the chalcogen atoms on one side. This TaS layer is honeycomb-like but not on the same plane, and a 2x2 Mn sublattice is stacked on the chalcogen side. From the TEM measurement and crystal structure analysis, the intercalant's position is aligned with transition metal atom in $2H_a$ TMD, while in $2H_c$ stacking it is also possible to be in the interstitial site. In the following model the 2x2 Mn are directly aligned with Ta (Fig. S2a). The unit cell has 4 Ta, 4 S and 1 Mn atoms and the Hamiltonian is a 9x9 Hermitian matrix when only s-orbitals are taken into consideration. Here we adopt the lattice gauge and the nearest neighbor hopping model can be written as

$$H = \begin{pmatrix} \varepsilon_{Ta} & 0 & 0 & 0 & t_1 & e^{-ik_1}t_1 & e^{-ik_2}t_1 & 0 & 0 \\ 0 & \varepsilon_{Ta} & 0 & 0 & t_1 & t_1 & 0 & e^{-ik_2}t_1 & 0 \\ 0 & 0 & \varepsilon_{Ta} & 0 & t_1 & 0 & t_1 & e^{-ik_1}t_1 & 0 \\ 0 & 0 & 0 & \varepsilon_{Ta} & 0 & t_1 & t_1 & t_1 & 0 \\ t_1 & t_1 & t_1 & 0 & \varepsilon_S & 0 & 0 & 0 & t_2 \\ e^{ik_1}t_1 & t_1 & 0 & t_1 & 0 & \varepsilon_S & 0 & 0 & e^{ik_1}t_2 \\ e^{ik_2}t_1 & 0 & t_1 & t_1 & 0 & 0 & \varepsilon_S & 0 & e^{ik_2}t_2 \\ 0 & e^{ik_2}t_1 & e^{ik_1}t_1 & t_1 & 0 & 0 & 0 & \varepsilon_S & 0 \\ 0 & 0 & 0 & 0 & t_2 & e^{-ik_1}t_2 & e^{-ik_2}t_2 & 0 & \varepsilon_{Mn} \end{pmatrix}$$

where $t_1$ and $t_2$ are the nearest neighbor hopping between Ta-S and Mn-S, $k_1$ and $k_2$ are phases with $k_i = \vec{k} \cdot \vec{r_i}$, and $\varepsilon$ denotes different elements' onsite potentials. Each Mn atom has a Ta symmetrically aligned with respect to the S plane and share the same hopping phases. The differences between these two sites are onsite potentials and hopping strengths. Therefore we can rewrite the Hamiltonian in a new basis by normalizing their states with a unitary transformation $H_{new} = U^{\dagger} H U$. Only sites of Mn and the aligned Ta are modified. Here the unitary matrix U and the new Hamiltonian take the form

$$U = \begin{pmatrix} \frac{t_1}{\sqrt{{t_1}^2 + {t_2}^2}} & 0 & \frac{t_2}{\sqrt{{t_1}^2 + {t_2}^2}} \\ 0 & I_{7\times 7} & 0 \\ -\frac{t_2}{\sqrt{{t_1}^2 + {t_2}^2}} & 0 & \frac{t_1}{\sqrt{{t_1}^2 + {t_2}^2}} \end{pmatrix}$$

$$H_{new} =$$

$$\begin{pmatrix} \frac{{t_1}^2 \varepsilon_{Ta} + {t_2}^2 \varepsilon_{Mn}}{{t_1}^2 + {t_2}^2} & 0 & 0 & 0 & \sqrt{{t_1}^2 + {t_2}^2} & e^{-ik_1}\sqrt{{t_1}^2 + {t_2}^2} & e^{-ik_2}\sqrt{{t_1}^2 + {t_2}^2} & 0 & \frac{t_1 t_2 (\varepsilon_{Mn} - \varepsilon_{Ta})}{{t_1}^2 + {t_2}^2} \\ 0 & \varepsilon_{Ta} & 0 & 0 & t_1 & t_1 & 0 & e^{-ik_2} t_1 & 0 \\ 0 & 0 & \varepsilon_{Ta} & 0 & t_1 & 0 & t_1 & e^{-ik_1} t_1 & 0 \\ 0 & 0 & 0 & \varepsilon_{Ta} & 0 & t_1 & t_1 & t_1 & 0 \\ \sqrt{{t_1}^2 + {t_2}^2} & t_1 & t_1 & 0 & \varepsilon_S & 0 & 0 & 0 & 0 \\ e^{ik_1}\sqrt{{t_1}^2 + {t_2}^2} & t_1 & 0 & t_1 & 0 & \varepsilon_S & 0 & 0 & 0 \\ e^{ik_2}\sqrt{{t_1}^2 + {t_2}^2} & 0 & t_1 & t_1 & 0 & 0 & \varepsilon_S & 0 & 0 \\ 0 & e^{ik_2} t_1 & e^{ik_1} t_1 & t_1 & 0 & 0 & 0 & \varepsilon_S & 0 \\ \frac{t_1 t_2 (\varepsilon_{Mn} - \varepsilon_{Ta})}{{t_1}^2 + {t_2}^2} & 0 & 0 & 0 & 0 & 0 & 0 & 0 & \frac{{t_1}^2 \varepsilon_{Mn} + {t_2}^2 \varepsilon_{Ta}}{{t_1}^2 + {t_2}^2} \end{pmatrix}$$

The new effective Hamiltonian replaces the original Ta site by a hybridized state $\frac{1}{t_2}|Ta> + \frac{1}{t_1}|Mn>$ with effective onsite potential $\frac{{t_1}^2 \varepsilon_{Ta} + {t_2}^2 \varepsilon_{Mn}}{{t_1}^2 + {t_2}^2}$ and hopping parameter $\sqrt{{t_1}^2 + {t_2}^2}$ to S, and the original Mn site by a destructive state $\frac{1}{t_1}|Ta> - \frac{1}{t_2}|Mn>$ with effective onsite potential $\frac{{t_1}^2 \varepsilon_{Mn} + {t_2}^2 \varepsilon_{Ta}}{{t_1}^2 + {t_2}^2}$ and hopping parameter $t_{eff} = \frac{t_1 t_2 (\varepsilon_{Mn} - \varepsilon_{Ta})}{{t_1}^2 + {t_2}^2}$ only to Ta. It is worth noting that the hopping between Mn and S is eliminated and the remaining hopping is confined between Mn and Ta without k-dispersion. If the onsite potentials of Mn and Ta are the same, the hopping between Mn and Ta becomes zero and a flat band $\varepsilon_{eff} = \frac{{t_1}^2 \varepsilon_{Mn} + {t_2}^2 \varepsilon_{Ta}}{{t_1}^2 + {t_2}^2} = \varepsilon_{Mn}$ emerges. The corresponding localized state is then $\frac{1}{t_1}|Ta> - \frac{1}{t_2}|Mn>$. The Ta and Mn have

opposite phases and cancel out on the S sites.

In more general cases, the Mn and Ta onsite potentials are different, the flat band starts to disperse slightly, and the localization is weakened with some wavefunction extension to S sites. To better understand how flat band evolves when the Mn onsite potential $\varepsilon_{Mn}$ changes, we calculate the characteristic polynomial of the new Hamiltonian $f_k(\lambda) = \det(H_{new} - \lambda I)$. Eigenvalues of the Hamiltonian are roots of $f_k(\lambda)$ where k means the solved $\lambda$ is momentum dependent. By expanding the determinant of $H_{new} - \lambda I$ along the last row, the result can be derived as

$$f_k(\lambda) = \left(\varepsilon_{eff} - \lambda\right) \cdot f_k^{TaS}(\lambda) - {t_{eff}}^2 \cdot f_k^{HV}(\lambda)$$

Here $f_k^{TaS}(\lambda)$ is the characteristic polynomial of 2x2 TaS Hamiltonian (which is an 8x8 Hermitian matrix) by deleting the last row and column (the Mn contributions) in $H_{new} - \lambda I$. The eigenvalue spectrum is almost the same as the original 2x2 TaS superlattice and small deviations come from the modified first effective Ta state (Fig. S2b). $f_k^{HV}(\lambda)$ is the characteristic polynomial of the 7x7 Hermitian principal submatrix by further deleting the first row and column (corresponding to the Ta aligned with Mn). The corresponding structure is a honeycomb lattice with periodic 2x2 vacancies (denote as HV - honeycomb vacancy lattice) as shown in Fig. S2c. The HV lattice consists of a Ta kagome sublattice and a defected S hexagonal sublattice. By solving the HV lattice Hamiltonian (Fig. S2d), we find that this HV sublattice contains three flat bands and the dispersionless constant energies are $\lambda_1 = \varepsilon_S, \lambda_{2,3} = \frac{\varepsilon_S + \varepsilon_{Ta}}{2} \mp \sqrt{t_1^2 + \left(\frac{\varepsilon_S - \varepsilon_{Ta}}{2}\right)^2}$, which are the S onsite potential and Ta-S bonding/antibonding energies, respectively. Fig. S2e shows those flat bands' corresponding localized states with the numbers showing the eigenfunction components' amplitudes and phases. Now we take a closer look at the whole system $f_k(\lambda)$. The dispersive terms only involve two characteristic polynomials. To eliminate the momentum dispersion, the coefficients should be zero. The first flat band possibility occurs when $t_{eff} = 0$ (onsite potentials of Mn and Ta are the same as discussed above) and then one root of $f_k(\lambda)$ is guaranteed to be a constant, $\lambda = \varepsilon_{eff}$. When the Mn/Ta onsite potentials are no longer the same and $t_{eff} \neq 0$, dispersionless solutions still exist when $\varepsilon_{eff} = \lambda_{1,2,3}$. It's not hard to find that while $\varepsilon_{eff} = \lambda_{1,2,3}$, we have $f_k(\lambda_i) = -{t_{eff}}^2 \cdot f_k^{HV}(\lambda_i) = 0$. $f_k(\lambda = \varepsilon_{eff})$ is zero and then $\varepsilon_{eff}$ is the root of $f_k(\lambda)$ which is a constant. This means flat bands can be preserved over a larger energy window more than $\varepsilon_{Mn} = \varepsilon_{Ta}$. The corresponding localized states are the same as Fig. S2e but with additional destructive states $\frac{1}{t_1}|Ta> - \frac{1}{t_2}|Mn>$ connected with S at the boundary. Fig. S2f gives the schematic illustration of different onsite potentials and energy levels. These together produce the dominant bands near the Fermi level. While $\varepsilon_{Mn}$ varies, the corresponding $\varepsilon_{eff}$ also varies (dashed line) and it can satisfy or get close to one of the four perfect flat band requirements (black lines), so one nearly flat or segmented flat band (due to band hybridization and avoided crossings) can emerge around the Fermi level. In all these conditions, the flat band position is set by $\varepsilon_{eff}$, which sits between the Mn and Ta onsite potentials. The $\lambda_{1,2,3}$ positions on the other hand mostly vary with the S onsite potential.

In summary, the destructive interference state $\frac{1}{t_1}|Ta> -\frac{1}{t_2}|Mn>$ contributes to the flat band formation. When their onsite potentials are the same or close, the dispersive hopping term $t_{eff} = \frac{t_1 t_2(\varepsilon_{Mn}-\varepsilon_{Ta})}{{t_1}^2+{t_2}^2} \leq \frac{\varepsilon_{Mn}-\varepsilon_{Ta}}{2}$ is negligible, and as a result Mn/Ta can cancel each other on S sites. When the onsite potential difference is getting larger, flat band is still possible when the effective onsite potential matches one of the flat bands of the HV sublattice, and the localization is extended to S sites. Overall, for typical materials, the orbitals of interest normally have their onsite potentials around the Fermi level, and a flat or nearly flat band could readily show up in a wide range of parameters.

## SII. Next nearest neighbor (NNN) hopping in tight binding model

When NNN hopping is taken into consideration, the Hamiltonian can be written as

$$H =$$

$$\begin{pmatrix}
\varepsilon_{Ta} & (1+e^{-ik_1})t_{Ta} & (1+e^{-ik_2})t_{Ta} & (e^{-ik_1}+e^{-ik_2})t_{Ta} & t_1 & e^{-ik_1}t_1 & e^{-ik_2}t_1 & 0 & t_{Mn} \\
(1+e^{ik_1})t_{Ta} & \varepsilon_{Ta} & \left(1+e^{i(k_1-k_2)}\right)t_{Ta} & (1+e^{-ik_2})t_{Ta} & t_1 & t_1 & 0 & e^{-ik_2}t_1 & 0 \\
(1+e^{ik_2})t_{Ta} & \left(1+e^{-i(k_1-k_2)}\right)t_{Ta} & \varepsilon_{Ta} & (1+e^{-ik_1})t_{Ta} & t_1 & 0 & t_1 & e^{-ik_1}t_1 & 0 \\
(e^{ik_1}+e^{ik_2})t_{Ta} & (1+e^{ik_2})t_{Ta} & (1+e^{ik_1})t_{Ta} & \varepsilon_{Ta} & 0 & t_1 & t_1 & t_1 & 0 \\
t_1 & t_1 & t_1 & 0 & \varepsilon_S & (1+e^{-ik_1})t_S & (1+e^{-ik_2})t_S & (e^{-ik_1}+e^{-ik_2})t_S & t_2 \\
e^{ik_1}t_1 & t_1 & 0 & t_1 & (1+e^{ik_1})t_S & \varepsilon_S & (1+e^{ik_1})t_S & (1+e^{-ik_2})t_S & e^{ik_1}t_2 \\
e^{ik_2}t_1 & 0 & t_1 & t_1 & (1+e^{ik_2})t_S & \left(1+e^{-i(k_1-k_2)}\right)t_S & \varepsilon_S & (1+e^{-ik_1})t_S & e^{ik_2}t_2 \\
0 & e^{ik_2}t_1 & e^{ik_1}t_1 & t_1 & (e^{ik_1}+e^{ik_2})t_S & (1+e^{ik_2})t_S & (1+e^{ik_1})t_S & \varepsilon_S & 0 \\
t_{Mn} & 0 & 0 & 0 & \mathrm{t2} & e^{-ik_1}t_2 & e^{-ik_2}t_2 & 0 & \varepsilon_{Mn}
\end{pmatrix}$$

where $t_{Ta}$, $t_S$ and $t_{Mn}$ are the NNN hopping parameters between Ta-Ta, S-S and Mn-Ta. The dilute intercalation makes sure that Mn-Mn interaction is negligible. The Hamiltonian can be divided into several blocks

$$H = \begin{pmatrix} H_{Ta} & T_{Ta-S} & T^{\dagger}_{Mn-Ta} \\ T^{\dagger}_{Ta-S} & H_S & T^{\dagger}_{Mn-S} \\ T_{Mn-Ta} & T_{Mn-S} & H_{Mn} \end{pmatrix}$$

where

$$H_{Ta} = \varepsilon_{Ta} I_{4\times 4} + t_{Ta} T_{NNN}$$

$$H_S = \varepsilon_S I_{4\times4} + t_S T_{NNN}$$

$$H_{Mn} = \varepsilon_{Mn}$$

$$T_{NNN} = \begin{pmatrix} 0 & 1+e^{-ik_1} & 1+e^{-ik_2} & e^{-ik_1}+e^{-ik_2} \\ 1+e^{ik_1} & 0 & 1+e^{i(k_1-k_2)} & 1+e^{-ik_2} \\ 1+e^{ik_2} & 1+e^{-i(k_1-k_2)} & 0 & 1+e^{-ik_1} \\ e^{ik_1}+e^{ik_2} & 1+e^{ik_2} & 1+e^{ik_1} & 0 \end{pmatrix}$$

$$T_{Ta-S} = t_1 \begin{pmatrix} 1 & e^{-ik_1} & e^{-ik_2} & 0 \\ 1 & 1 & 0 & e^{-ik_2} \\ 1 & 0 & 1 & e^{-ik_1} \\ 0 & 1 & 1 & 1 \end{pmatrix}$$

$$T_{Mn-S} = t_2 (1 \quad e^{-ik_1} \quad e^{-ik_2} \quad 0)$$

$$T_{Mn-Ta} = t_{Mn} (1 \quad 0 \quad 0 \quad 0)$$

Fig. S3 gives example band structures of different tight binding parameters as well as NNN hopping. The line width and opacity indicate the contribution after band unfolding from reduced supercell Brillouin zone into primitive Brillouin zone, by considering the Bloch phases of identical atoms in the supercell and calculating the spectral weights of eigenvalues at each k-point[1]. The overall nearly flat features can be observed in these band structures. The deviation from perfect flatness with dispersive and segmented bands come from the NNN hopping and band hybridization with gap opening at avoided band crossings.

**SIII. Set-up of polarization dependent ARPES**

According to the selection rule in linear polarized photoemission, bands having the same parity with respect to the mirror plane can be selectively resolved. Fig. S8a gives the schematic illustration of our ARPES experimental geometry. In the ARPES set-up employed in our measurements, the mirror plane is defined by the photon incident direction and the normal of sample surface, as well as horizontal analyzer slit. When the photon is linear horizontal (linear vertical) polarized, the electric field vector lies in (orthogonal to) the mirror plane, then only orbitals that have even (odd) parity can be resolved. The parity symmetries of d and p orbitals are summarized in Fig. S8b. The selection rule of polarization dependence is well revealed in both the ARPES measurements and DFT calculations of the host $TaS_2$ band structures. In Fig. S8c the projected band structures of dominate Ta d orbitals and S p orbitals in linear polarizations are shown. The contribution in linear horizontal polarized ARPES measurement mainly comes from Ta $d_{z2}/d_{xz}$ and S $p_z/p_x$ orbitals, while in linear vertical polarization comes from Ta $d_{yz}$ and S $p_y$ orbitals.

**SIV. The localized states of flat bands in tight binding model**

In SI we have discussed the tight binding model where intercalants are aligned with transition metal atoms (which is the case in intercalated $2H_a$-TMDs). When the Mn intercalants are not aligned with Ta but interstitial with both Ta and S (correspond to intercalation in $2H_c$-TMDs), the tight binding structure is almost the same as above, except the position of intercalants. The Hamiltonian with nearest neighbor hopping then can be written as

$$H=\begin{pmatrix} \varepsilon_{Ta} & 0 & 0 & 0 & t_1 & e^{-ik_1}t_1 & e^{-ik_2}t_1 & 0 & 0 \\ 0 & \varepsilon_{Ta} & 0 & 0 & t_1 & t_1 & 0 & e^{-ik_2}t_1 & 0 \\ 0 & 0 & \varepsilon_{Ta} & 0 & t_1 & 0 & t_1 & e^{-ik_1}t_1 & 0 \\ 0 & 0 & 0 & \varepsilon_{Ta} & 0 & t_1 & t_1 & t_1 & 0 \\ t_1 & t_1 & t_1 & 0 & \varepsilon_S & 0 & 0 & 0 & t_2 \\ e^{ik_1}t_1 & t_1 & 0 & t_1 & 0 & \varepsilon_S & 0 & 0 & t_2 \\ e^{ik_2}t_1 & 0 & t_1 & t_1 & 0 & 0 & \varepsilon_S & 0 & t_2 \\ 0 & e^{ik_2}t_1 & e^{ik_1}t_1 & t_1 & 0 & 0 & 0 & \varepsilon_S & 0 \\ 0 & 0 & 0 & 0 & t_2 & t_2 & t_2 & 0 & \varepsilon_{Mn} \end{pmatrix}$$

The hopping phases from Mn are no longer the same as any Ta. By solving the eigenequation of Hamiltonian matrix, we find that the spectra of eigenvalues are the same as before and the difference comes from the eigenvectors. When Mn onsite potential is equal to that of Ta, a flat band located at $\varepsilon_{Ta}$ is observed but the localization is quite different. The eigenstate of this flat band is

$$v=\begin{pmatrix} t_2\left(1+e^{ik_1}+e^{ik_2}\right) \\ t_2\left(-1+e^{ik_2}-e^{2ik_1}+e^{ik_1+ik_2}\right) \\ t_2\left(-1+e^{ik_1}-e^{2ik_2}+e^{ik_1+ik_2}\right) \\ t_2\left(e^{ik_1}+e^{ik_2}-e^{2ik_1}-e^{2ik_2}\right) \\ 0 \\ 0 \\ 0 \\ 0 \\ t_1\left(1-2e^{ik_1}-2e^{ik_2}+e^{2ik_1}+e^{2ik_2}-2e^{ik_1+ik_2}\right) \end{pmatrix}$$

in the form of finite sum of the Bloch phases. By applying Fourier transformation on the eigenvector[2], we can get the localized geometry of flat band in real space. The electronic localization is confined in a large triangle involving 6 Mn atoms and 15 Ta atoms as shown in the main text (Fig. 4b). The wavefunction on S sites are cancelled everywhere along the perimeter. Similarly, when there is onsite potential difference between Mn and Ta, the localization starts to extend to S sites. Fig. S10a gives an example of the flat band's localized state when onsite potentials of Mn and Ta are different. Here the effective onsite potential $\varepsilon_{eff}$ of Mn and Ta is set to be equal to that of S. Compared with the localization shown in Fig. 4b, the triangular pocket involves addition S atoms inside with nonzero amplitudes. These S sites cancel each other on nearby Ta, and contribute to the electronic localization together with Mn/Ta.

We have discussed the real space localization of flat bands in the most common ordered and dilute supercell intercalation positions. For other complicated intercalation cases, the flat bands can be achieved by the superposition of individual localized pockets. For example, when there are 2 Mn atoms (Fig. S10b) in 2x2 interstitial intercalation (works for aligned stacking as well), the Hamiltonian can be written as

$$H =$$

$$\begin{pmatrix} \varepsilon_{Ta} & 0 & 0 & 0 & t_1 & e^{-ik_1}t_1 & e^{-ik_2}t_1 & 0 & 0 & 0 \\ 0 & \varepsilon_{Ta} & 0 & 0 & t_1 & t_1 & 0 & e^{-ik_2}t_1 & 0 & 0 \\ 0 & 0 & \varepsilon_{Ta} & 0 & t_1 & 0 & t_1 & e^{-ik_1}t_1 & 0 & 0 \\ 0 & 0 & 0 & \varepsilon_{Ta} & 0 & t_1 & t_1 & t_1 & 0 & 0 \\ t_1 & t_1 & t_1 & 0 & \varepsilon_S & 0 & 0 & 0 & t_2 & e^{-ik_1}t_2 \\ e^{ik_1}t_1 & t_1 & 0 & t_1 & 0 & \varepsilon_S & 0 & 0 & t_2 & t_2 \\ e^{ik_2}t_1 & 0 & t_1 & t_1 & 0 & 0 & \varepsilon_S & 0 & t_2 & 0 \\ 0 & e^{ik_2}t_1 & e^{ik_1}t_1 & t_1 & 0 & 0 & 0 & \varepsilon_S & 0 & t_2 \\ 0 & 0 & 0 & 0 & t_2 & t_2 & t_2 & 0 & \varepsilon_{Mn} & 0 \\ 0 & 0 & 0 & 0 & e^{ik_1}t_2 & t_2 & 0 & t_2 & 0 & \varepsilon_{Mn} \end{pmatrix}$$

Here the last two rows and columns are hopping of $Mn_A$ and $Mn_B$ sites. Let $H_A$ and $v_A$ be the Hamiltonian and localized eigenstate by deleting $Mn_B$, then $H$ can be expressed as

$$H = \begin{pmatrix} H_A & P^\dagger \\ P & \varepsilon_{Mn} \end{pmatrix}$$

where $P = (0 \quad 0 \quad 0 \quad 0 \quad e^{ik_1}t_2 \quad t_2 \quad 0 \quad t_2 \quad 0)$ is the hopping between $Mn_B$-Ta, $Mn_B$-S and $Mn_B$- $Mn_A$, with only the $Mn_B$-S hopping allowed. It's not hard to find that the eigenvector of $H$ can be obtained by adding 0 at $Mn_B$ site in $v_A$ (localization is shown in Fig. S10c)

$$H\, v_A' = \begin{pmatrix} H_A & P^\dagger \\ P & \varepsilon_{Mn} \end{pmatrix} \begin{pmatrix} v_A \\ 0 \end{pmatrix} = \begin{pmatrix} H_A\, v_A \\ P\, v_A \end{pmatrix} = \begin{pmatrix} H_A\, v_A \\ 0 \end{pmatrix} = E\, v_A'$$

This is because $v_A$ is zero at S sites due to the destructive cancellation while $P$ is only non-zero at S sites connected to $Mn_B$. As a result, $P\, v_A$ is equal to zero. On the other hand, because all S sites are zero due to destructive interference, they have no hopping contributions to $Mn_B$ sites, so we can add 0 at $Mn_B$ site in $v_A$ as the eigenstate and localization can be kept in this mixed structure. This gives one of the eigenvectors of $H$, and the localization corresponds to the case of single intercalant atom in the unit cell. These are still valid when $Mn_A$ and $Mn_B$ sites are swapped, which means $v_B'$ constructed by the same way (add 0 at $Mn_A$ site in $v_B$) is the eigenvector as well. These two eigenstates have the same eigenvalues, therefore any superposition of the two individual localized pockets $\alpha\, v_A' + \beta\, v_B'$ is still the eigenvector of $H$ and the flat bands are doubly degenerate.

When $v_A$ has finite non-zero values at S sites like Fig. S10a shows, corresponding to the $\varepsilon_{eff} = \varepsilon_S$ solution, the equality $P\, v_A = 0$ still holds. We can view this localization more clearly when we put $Mn_B$ at the interstitial center of any hexagon left in the structure shown in Fig. S10a. The sum of hopping from neighboring S is always zero, in agreement with the equality $P\, v_A = 0$. This makes sure that the individual localized state of each intercalant can still keep localization with additional intercalant in the unit cell. When theses intercalants are identical (having same tight binding parameters), the eigenvector is the arbitrary superposition of each individual localized state which is a consequence of flat bands degeneracy. When they are different atom species but satisfy their own dispersionless requirements, the degeneracy no longer exists and there will be multiple flat bands.

Another case is that the intercalation has supercells with different sizes mixed. We can denote these supercells as A and B sublattices (similar to the description of single intercalant atom A and B above), then the same method can be applied. For example, given an intercalation with 2x2 and $\sqrt{7} \times \sqrt{7}$ supercells mixed, we can find a lager common supercell that can enclose both types of intercalants (a 14x14 supercell here), and then construct the supercell Hamiltonian. Because flat bands still stay dispersionless after supercell band folding, the Hamiltonian can be divided into A and B blocks following the same way:

$$H = \begin{pmatrix} H_A & P^\dagger \\ P & \varepsilon_B \end{pmatrix}$$

where $H_A$ is the Hamiltonian with only one type intercalation, $\varepsilon_B$ is the hopping matrix among sublattice B and is diagonal with onsite potentials only due to dilute intercalation, and $P$ is the hopping between sublattice B with S (zero with Ta and sublattice A). Again the equality $P\, v_A = 0$ is valid because all S sites are zero due to destructive interference, or the nonzero S wavefunctions cancel each other on neighboring sites, therefore $v'_{A(B)}$ with zero at sublattice B(A) sites in $v_{A(B)}$ is the eigenvector of the flat band. Here the individual localized states are the triangular pockets summarized in main text of supercells with different sizes. As a result, real space localization can be achieved with degenerate or multiple flat bands depending on the intercalant species.

**SV. Generalized tight binding model and the kz dependence of the flatness**

Here we generalize tight binding model to the real structures of monolayer and bulk intercalated TMDs respectively. The monolayer $TaS_2$ structure with a layer of 2x2 Mn on top is shown in Fig. S11a. Compared with the simplified TaS model discussed above, there is an additional S layer stacked on the other side of Ta and aligned with the previous S layer. The Hamiltonian then has one more block and is a 13x13 Hermitian matrix

$$H = \begin{pmatrix} H_{Ta} & T_{Ta-S} & T_{Ta-S} & T_{Mn-Ta}^{\dagger} \\ T_{Ta-S}^{\dagger} & H_{S(bottom)} & T_{S-S} & 0 \\ T_{Ta-S}^{\dagger} & T_{S-S}^{\dagger} & H_{S(top)} & T_{Mn-S}^{\dagger} \\ T_{Mn-Ta} & 0 & T_{Mn-S} & H_{Mn} \end{pmatrix}$$

where $T_{S-S} = t_s I_{4\times4}$ is the interaction between top and bottom S layers. Because of the absence of Mn layer on the other side to cancel out on S, the strictly flat band disappears and starts to disperse slightly instead. Fig. S11a gives the example band structure with same parameters in Fig. S3d. The flat band located at $\varepsilon_{Ta}$ becomes segmented due to avoided band crossing and hybridization.

The bulk structure has two $TaS_2$ layers with a relative 180-degrees rotation. The Hamiltonian can be written as

$$H = \begin{pmatrix} H_{mono} & 0 & T_{bottom}^{\dagger} \\ 0 & H_{mono}^{*} & T_{top}^{\dagger} \\ T_{bottom} & T_{top} & H_{Mn} \end{pmatrix}$$

where $H_{mono}$ is the Hamiltonian of bottom $TaS_2$ monolayer and takes the form

$$H_{mono} = \begin{pmatrix} H_{Ta} & T_{Ta-S} & T_{Ta-S} \\ T_{Ta-S}^{\dagger} & H_S & T_{S-S} \\ T_{Ta-S}^{\dagger} & T_{S-S}^{\dagger} & H_S \end{pmatrix}$$

. $H_{mono}^{*}$ is the Hamiltonian of top $TaS_2$ monolayer and is the conjugate of $H_{mono}$, because of the relative rotation and opposite lattice vectors. $H_{Mn}$, $T_{top}$ and $T_{bottom}$ are Mn onsite potentials and hopping with adjacent $TaS_2$ layers and they take the form

$$H_{Mn} = \begin{pmatrix} \varepsilon_{Mn(bottom)} & 0 \\ 0 & \varepsilon_{Mn(top)} \end{pmatrix}$$

$$T_{bottom} = \begin{pmatrix} t_{Mn} & 0 & 0 & 0 & t_2 & e^{-ik_1}t_2 & e^{-ik_2}t_2 & 0 & 0 & 0 & 0 & 0 \\ t_{Mn} & 0 & 0 & 0 & 0 & 0 & 0 & t_2 & e^{-ik_1}t_2 & e^{-ik_2}t_2 & 0 \end{pmatrix}$$

$$T_{top} = \begin{pmatrix} e^{-i(k_1+k_2+k_3)}t_{Mn} & 0 & 0 & 0 & 0 & 0 & 0 & 0 & e^{-i(k_1+k_2+k_3)}t_2 & e^{-i(k_2+k_3)}t_2 & e^{-i(k_1+k_3)}t_2 & 0 \\ e^{-i(k_1+k_2)}t_{Mn} & 0 & 0 & 0 & e^{-i(k_1+k_2)}t_2 & e^{-ik_2}t_2 & e^{-ik_1}t_2 & 0 & 0 & 0 & 0 & 0 \end{pmatrix}$$

In this model the interlayer hopping is realized by Mn and the direct interaction between $TaS_2$ layers is neglected. The band structure is given in Fig. S11b. The bulk 3D structure contains a repeating cell with "Mn-$S_{top}$-Ta-$S_{bottom}$-Mn-…" which has an interlayer hopping. To have destructive interference on S, Mn and Ta will have alternating phases along the out-of-plane direction. Although there are flat bands at $k_z = 0$ plane, but the localization is two dimensional and extended along out-of-plane direction. Therefore they have $k_z$ dispersion and become slightly dispersive when interlayer interaction is taken into consideration (Fig. S12). This is consistent with DFT calculations (Fig. S13) that one of the flat bands at -1.23 eV can get dispersive to some extent at different $k_z$, which is a consequence of the delocalization along the out-of-plane direction.

The Hamiltonian of T phase TMD in main text is constructed by the same way, and all the tight binding parameters used in Fig. 4 and Fig. S14 are $\varepsilon_{Ta} = 0$eV, $\varepsilon_{S} = -6$eV, $\varepsilon_{Mn} = -2$eV, $t_1 = -5$eV, $t_2 = -3$eV, $t_{Ta} = -3$eV, $t_S = -2$eV and $t_{Mn} = -2$eV.

## Supplementary Table:

| EDC | Cut 1 | | Cut 2 | | Cut 3 | | Cut 4 | |
|---|---|---|---|---|---|---|---|---|
| Momentum (Å$^{-1}$) | $\bar{\Gamma}$ | $\bar{\mathrm{M}}$ | $\bar{\Gamma}$ | $\bar{K}$ | 0.0943785 | $\bar{K}$ | 0.093 | 1.019 |
| Amplitude $A$ | 90.600 | 936.799 | 3308.56 | 2639.04 | 765.583 | 772.457 | 1130.3 | 4423.91 |
| Position $E_0$ (eV) | -1.174 | -1.265 | -1.213 | -1.220 | -1.193 | -1.205 | -1.214 | -1.254 |
| FWHM $\Gamma$ (eV) | 0.448 | 0.938 | 0.955 | 0.830 | 0.889 | 0.956 | 0.653 | 0.922 |
| $b_0$ | 1220.18 | -451.861 | 5336.47 | 5793.19 | 1890.5 | 1769.46 | 2489.29 | 5853.67 |
| $b_1$ (eV$^{-1}$) | 2510.94 | -2063.22 | 10036.9 | 10152.8 | 3994.2 | 4075.93 | 2793.03 | 7689.96 |
| $b_2$ (eV$^{-2}$) | 1595.09 | -2968.63 | 2081.86 | 1915.49 | 2015.54 | 2398.32 | -1296.33 | -2496.84 |
| $b_3$ (eV$^{-3}$) | 234.36 | -1139.09 | -1274.14 | -1309.38 | 157.148 | 358.555 | -1492.33 | -2940.04 |

Table. S1: Fitting parameters of the flat band peaks in EDCs at high symmetry points.

## Supplementary Figures:

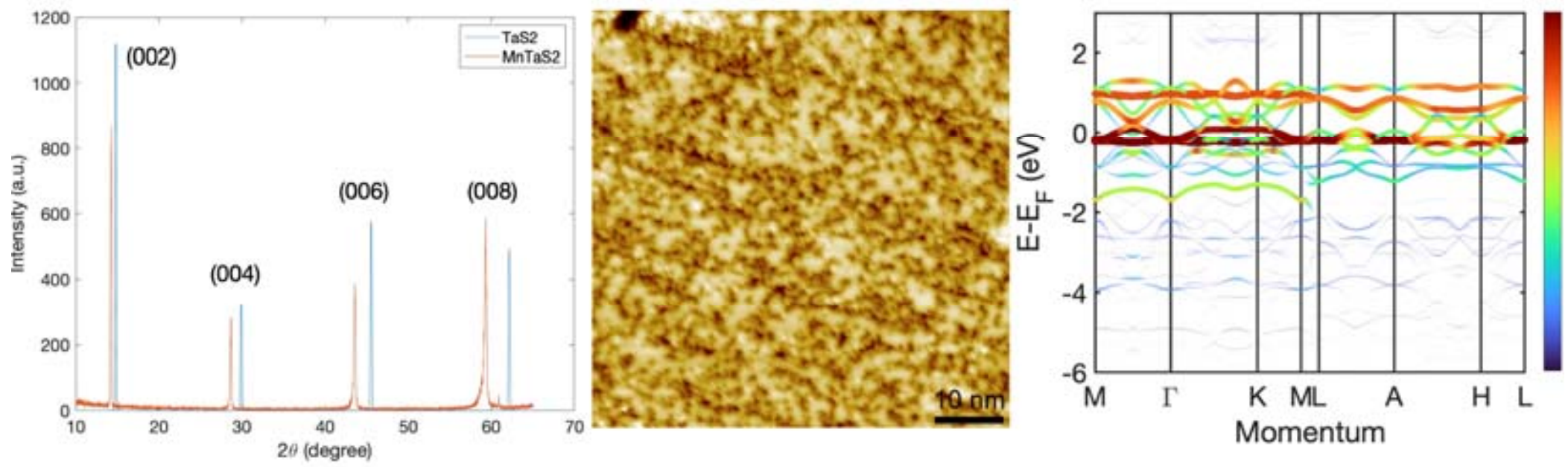

Fig. S1: Crystal characterization and DFT band structure calculation. **a**, XRD measurements of TaS2 and Mn1/4TaS2 single crystals. Calculated out-of-plane lattice constants are 12.46 Å for $Mn_{1/4}TaS_2$ and 11.94 Å for $TaS_2$, respectively. The shift to smaller angles is a direct consequence of the larger interlayer spacing with extra atoms. **b**, STM atomic resolution on Mn-terminated surface ($V_{bias}$ = 1 V, $I_{set}$ = 0.5 A). Scale bar is 10 nm. **c**, DFT band structure of $Mn_{1/4}TaS_2$ without spin polarization. Line width and color represent Mn weight projection.

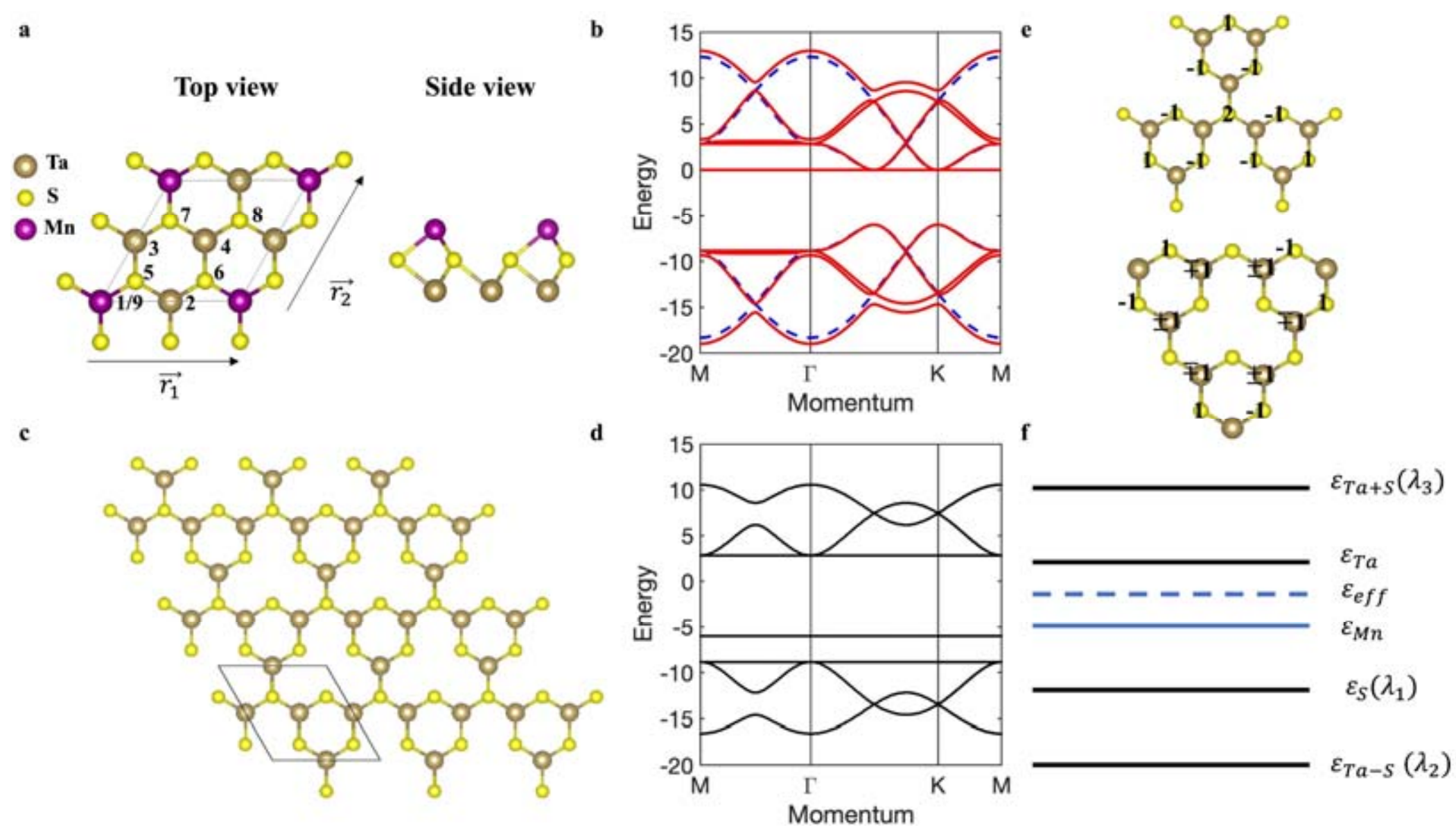

Fig. S2: Tight binding model for intercalated TMD. **a**, Side and top view of the simplified tight binding

structure. $\vec{r_1}$ and $\vec{r_2}$ are unit cell lattice vectors used for the hopping phase, and numbers near atoms indicate their positions (rows and columns) in the Hamiltonian. **b**, Band structures of 2x2 TaS (blue dashed line) and intercalated $Mn_{1/4}TaS$ (red solid line). **c**, Structure of honeycomb lattice with periodic 2x2 vacancies (HV lattice). The two sites in the unit cell can be two different atoms. **d**, Band structure of the HV lattice. **e**, Localized states of flat bands in **d**. The upper one corresponds to the solution $\lambda_1(\varepsilon_S)$ and the lower one corresponds to the solution $\lambda_{2,3}$. **f**, Schematic illustration of different onsite potentials and energy levels. The tight binding parameters used here and in Fig. 1d are $onsite = (\varepsilon_{Ta}, \varepsilon_S, \varepsilon_{Mn}) = (0, -6, 0)$ eV and $t = (t_1, t_2) = (-5, -3)$ eV.

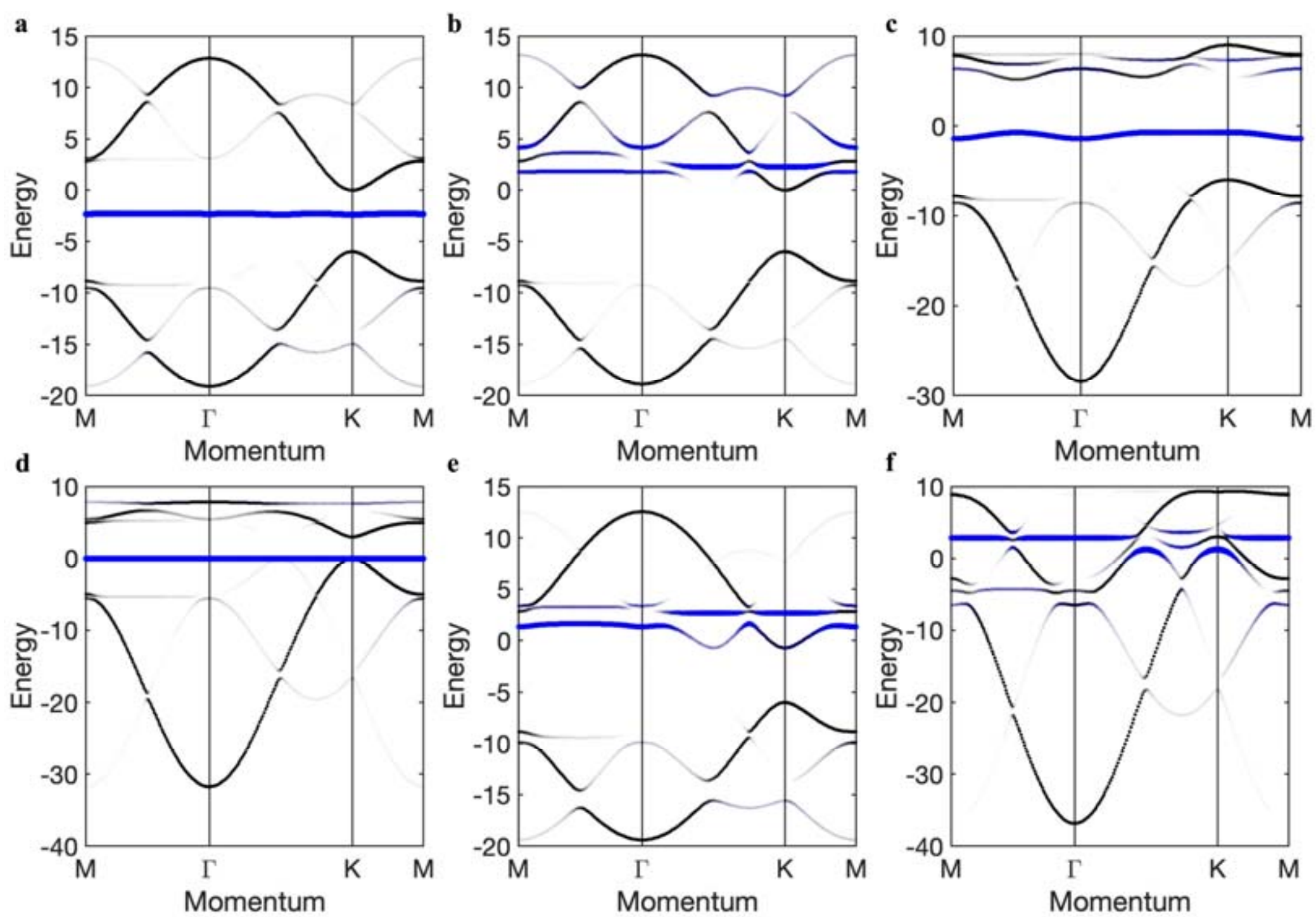

Fig. S3: Band structures with different tight binding parameters. **a**,**b**, Nearly flat bands when the onsite potential of Mn is different from that of Ta. The line width and opacity indicate the spectral weight after band unfolding, and the blue curves represent Mn flat bands. Parameters used are **a** $onsite = (0, -6, -3)$ eV and **b** $onsite = (0, -6, 3)$ eV with hopping $t = (-5, -3)$ eV. **c**-**f**, Nearly flat bands

when NNN hopping is taken into consideration. The NNN hopping parameters $t_{NNN} = (t_{Ta}, t_S, t_{Mn})$ are **c** $t_{NNN} = (-3, 0, 0)$ eV, **d** $t_{NNN} = (0, -3, 0)$ eV, **e** $t_{NNN} = (0, 0, -3)$ eV and **f** $t_{NNN} = (-3, -3, -3)$ eV.

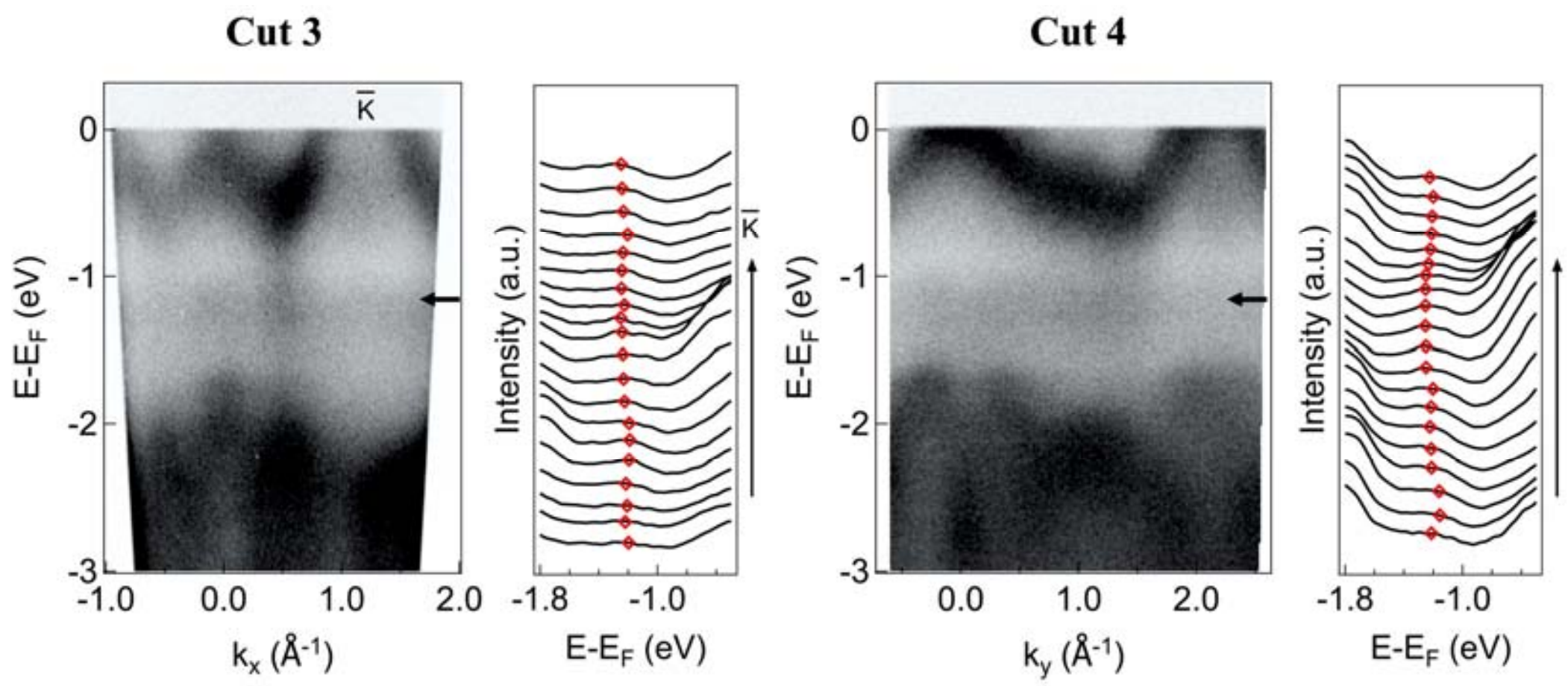

Fig. S4: ARPES spectra taken across high-symmetry points with corresponding EDCs. The directions in (left) Cut 3 and (right) Cut 4 are shown in Fig. 2a. Black arrows mark the flat band energy positions, and red rhombus dots track the flat band peaks in EDCs.

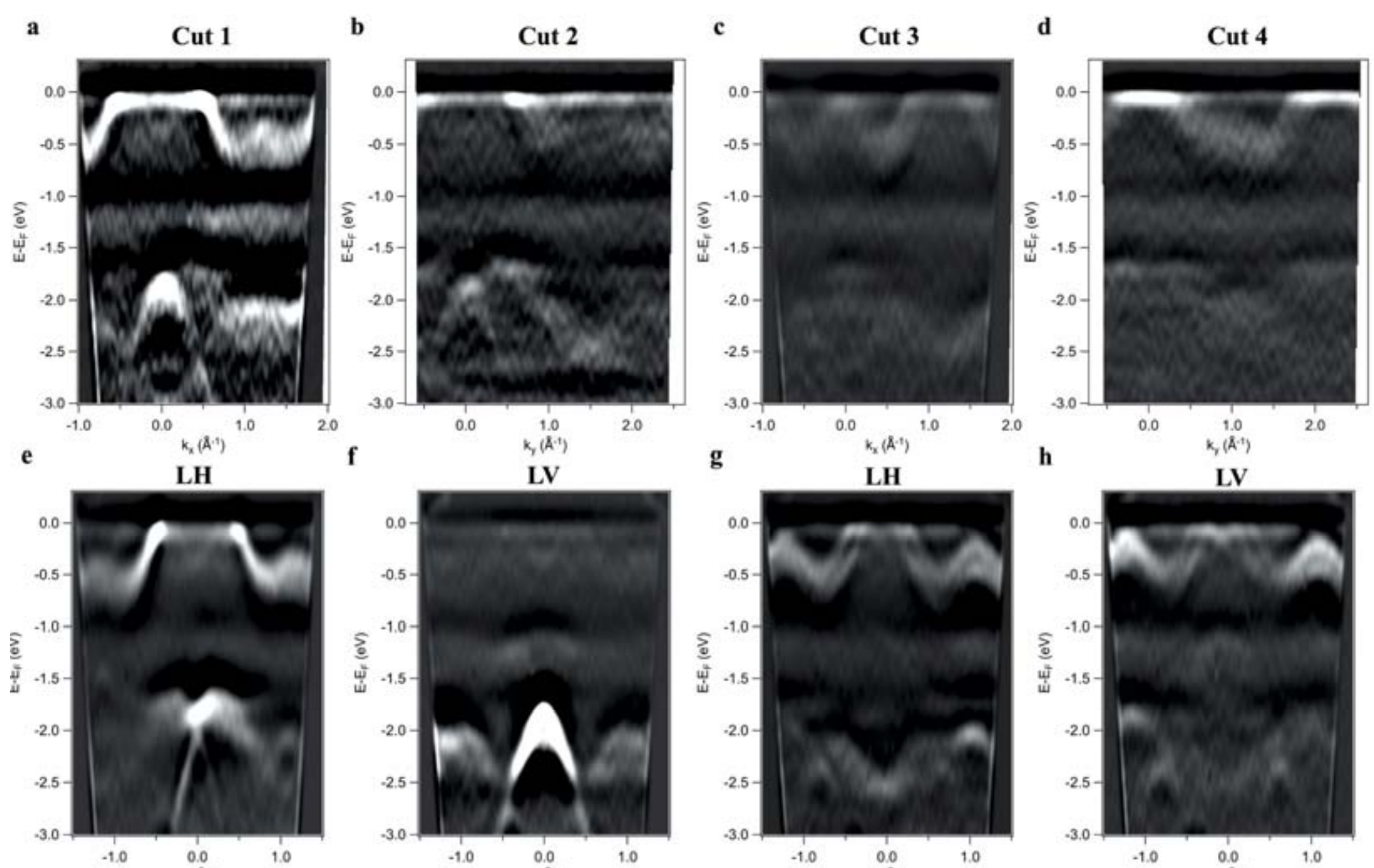

Fig. S5: Second derivative plots of ARPES spectra. **a-d**, Plots of Cut 1-4 shown in Fig. 2b and 2c, and Fig. S4. **e**,**f**, Plots of spectra along $\bar{\mathrm{M}}-\bar{\Gamma}-\bar{\mathrm{M}}$ direction under LH/LV polarizations shown in Fig. 3a. **g**,**h**, Plots of spectra along $\bar{\mathrm{K}}'-\bar{\mathrm{K}}-\bar{\mathrm{K}}'$ direction under LH/LV polarizations shown in Fig. S6. All of them show a clear flat band.

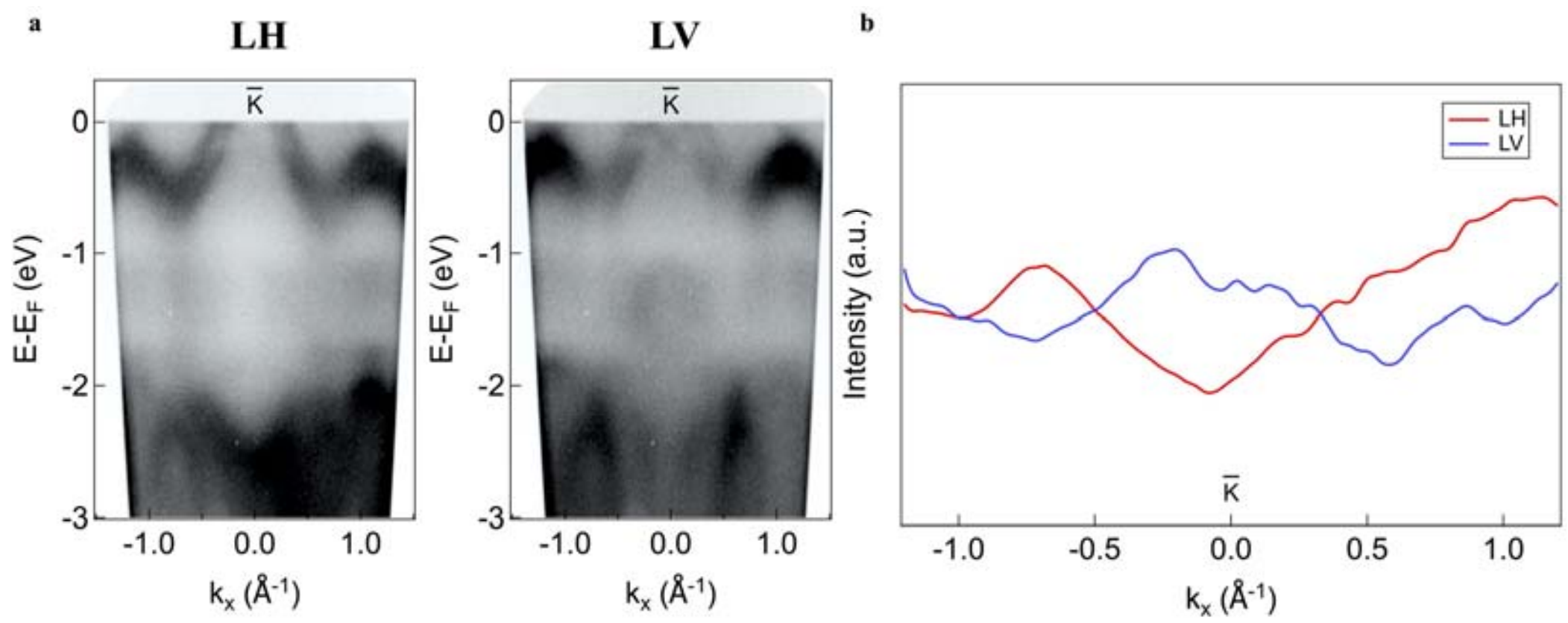

Fig. S6: Polarization dependent ARPES measurements across K point. **a**, ARPES spectra measured along $\bar{\mathrm{K}}'-\bar{\mathrm{K}}-\bar{\mathrm{K}}'$ direction with (left) LH and (right) LV polarizations. **b**, MDCs at the flat band position (integrated over an energy range of 40 meV around -1.23 eV) in (red) LH and (blue) LV polarizations.

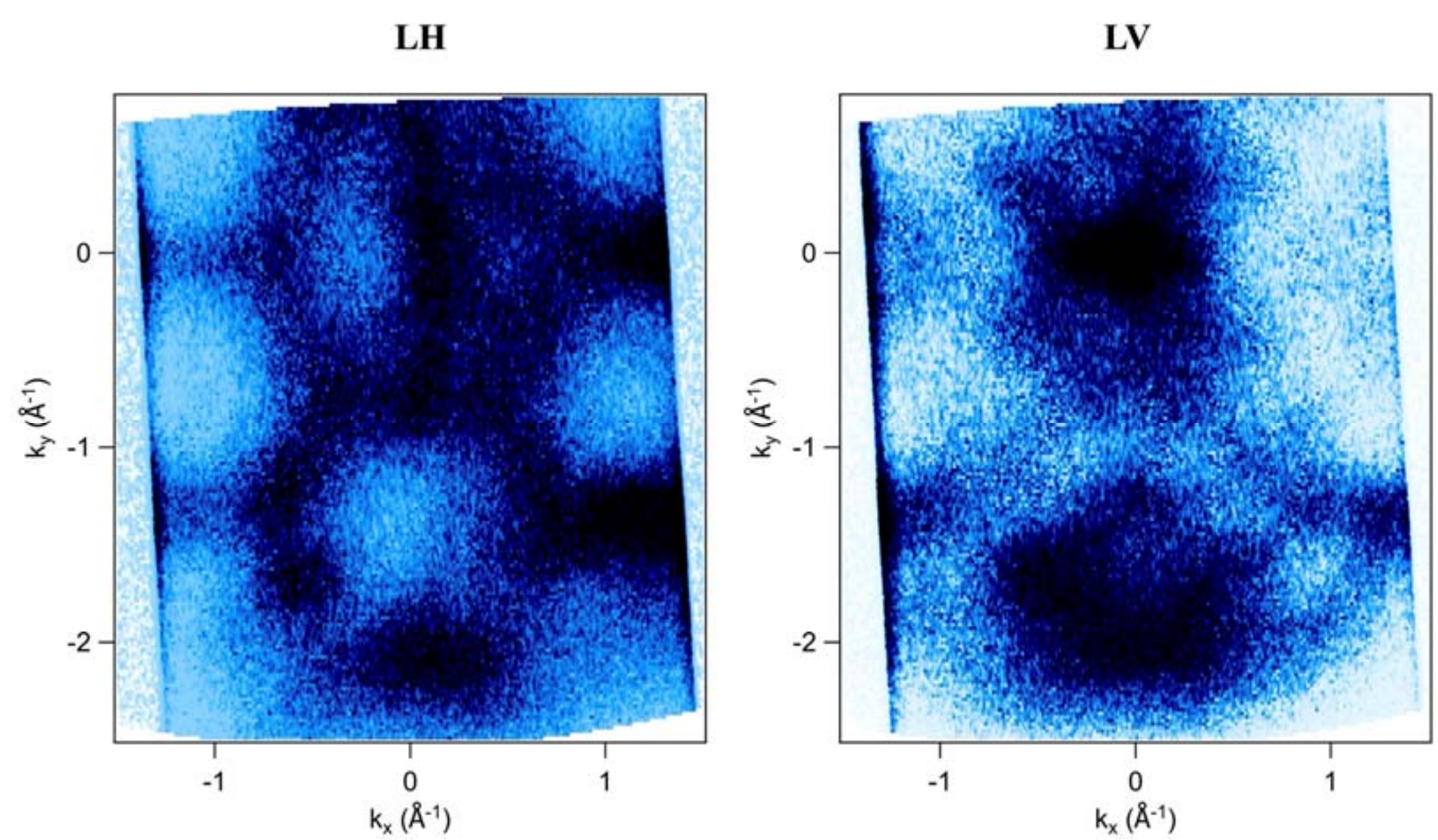

Fig. S7: Constant energy contour mapping at -1.23 eV where flat band is observed under linear horizontal (LH, left) and linear vertical (LV, right) polarizations.

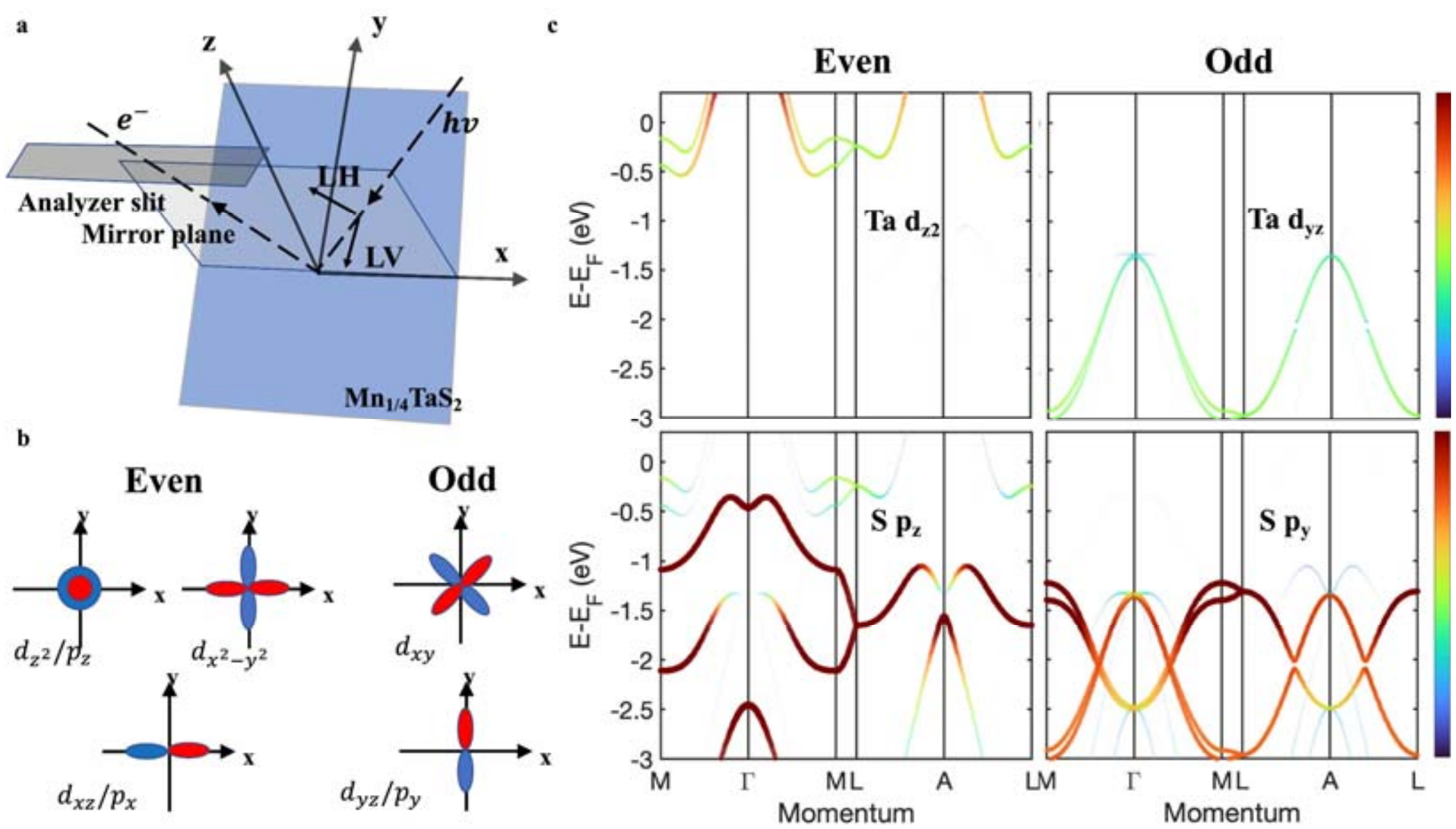

Fig. S8: Set-up of polarization dependent ARPES. **a**, Schematic illustration of ARPES experimental set-up. **b**, Parity symmetries of each d and p orbitals. The $d_{xz}$ ($d_{yz}$) orbital is equivalent to $p_x$ ($p_y$) orbital. **c**, Dominant orbital projected band structures of $TaS_2$. The line width and color depth represent projected orbital contributions in LH/LV polarizations.

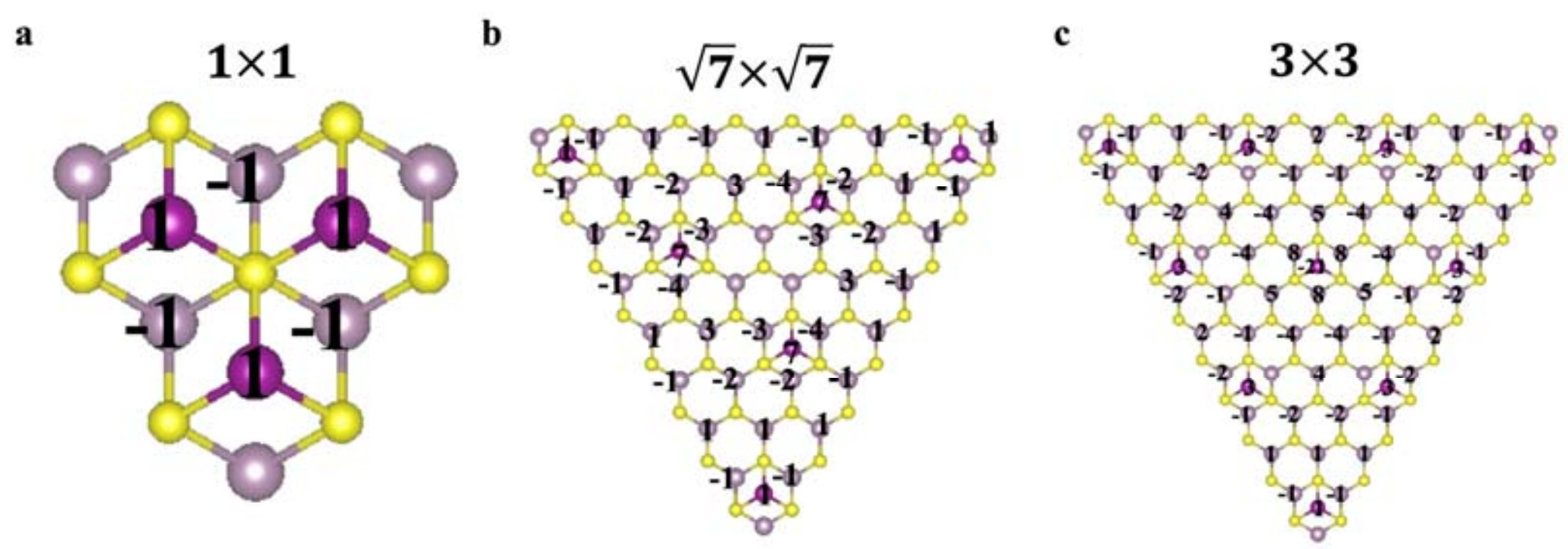

Fig. S9: Localized states of flat bands in **a** 1×1, **b** $\sqrt{7}\times\sqrt{7}$ and **c** 3×3 supercell intercalated structures.

The amplitude/phase of each atom is labeled

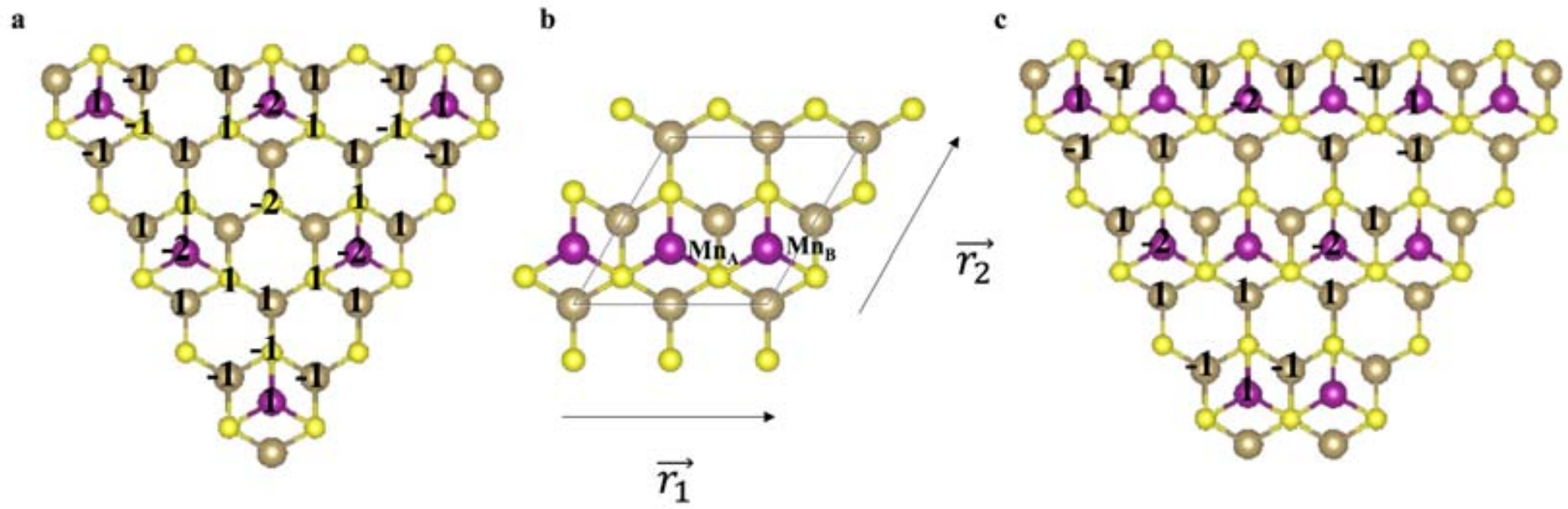

Fig. S10: Localization of flat bands in other intercalation cases. **a**, The localized state of interstitial intercalation ($2H_c$-TMD) with amplitude/phase of each atom labeled, corresponding to $\varepsilon_{eff} = \varepsilon_S$ solution. **b**, Tight bind structure of the interstitial intercalation with multiple intercalants (labeled by $Mn_A$ and $Mn_B$) in the unit cell. **c**, The localized state (same as single intercalant case) of the structure shown in **b**.

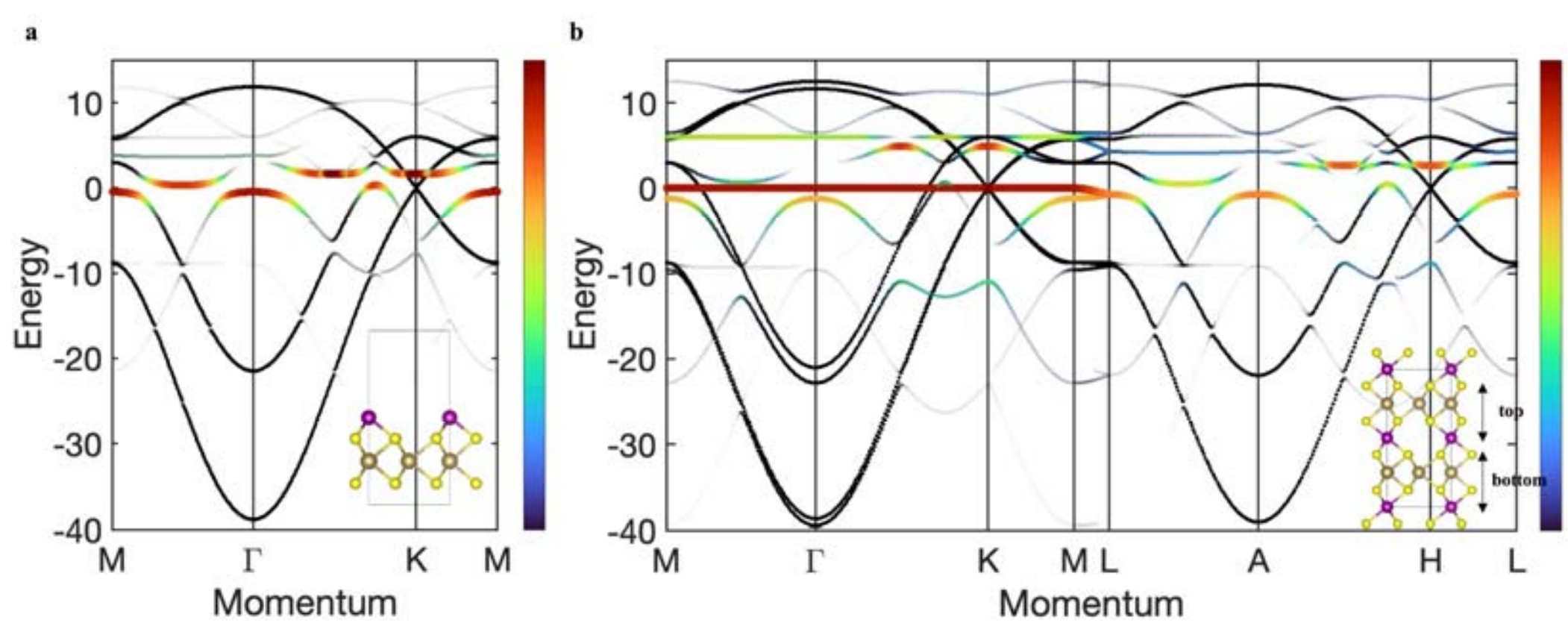

Fig. S11: Band structures of **a** monolayer and **b** bulk intercalated TMD. Insets are the structures used for tight binding modeling, and the bottom/top layer positions are indicated. Mn flat bands are marked by the yellowish color. The parameters used here are $onsite = (0, -6, 0)$ eV, $t = (-5, -3)$ eV and $t_{NNN} = (0, -3, 0)$ eV.

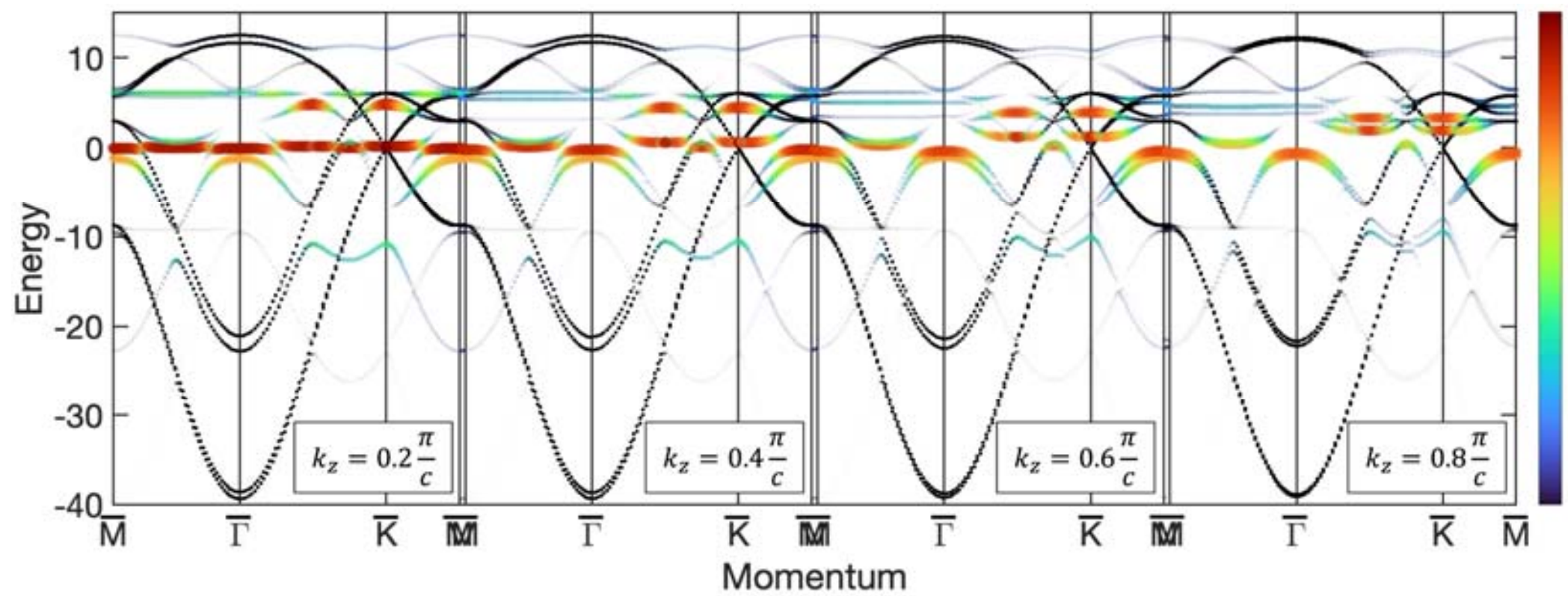

Fig. S12: Tight binding band structures of bulk intercalated TMD. The momentum $k_z$ is given for each band structure. Mn flat bands are marked by the yellowish color. The parameters used here are $onsite = (0, -6, 0)$ eV, $t = (-5, -3)$ eV and $t_{NNN} = (0, -3, 0)$ eV.

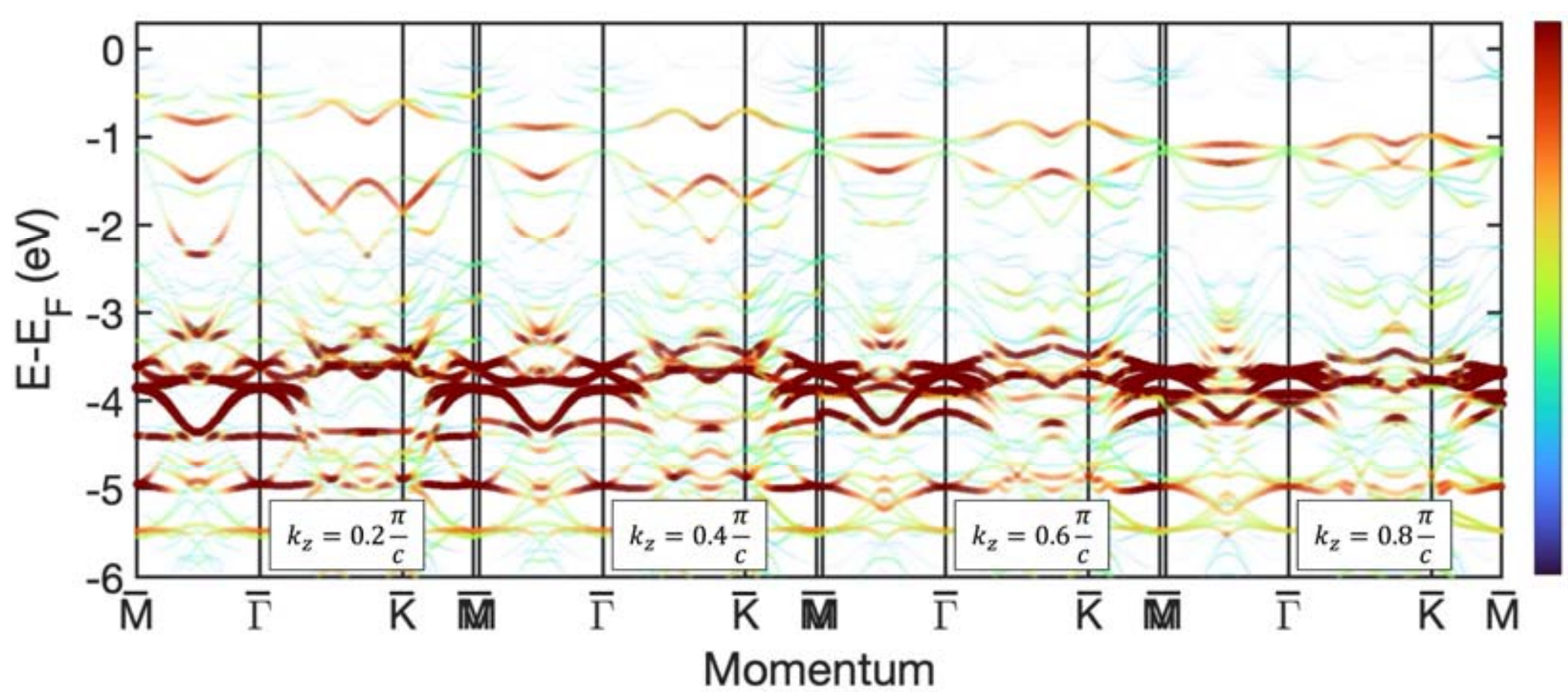

Fig. S13: DFT band structures of $Mn_{1/4}TaS_2$ at different $k_z$. The band structure with spin down component below the Fermi level is given. The line width and color represent flat band weight projections of Mn.

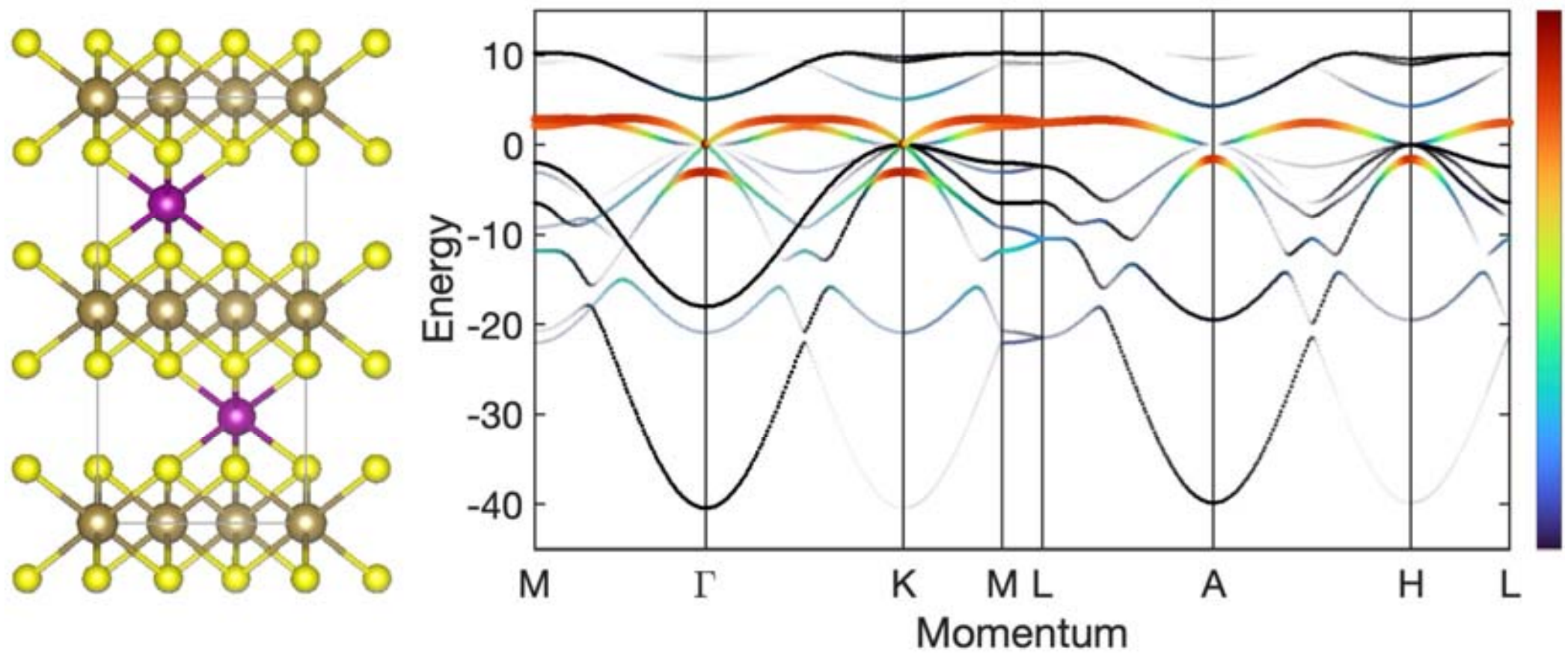

Fig. S14: Tight binding model and band structure of $\sqrt{3} \times \sqrt{3}$ T-TMD. The unit cell has two T phase layers due to intercalants misalignment. Line width and color represent Mn weight projection to the flat band. The parameters used here are $onsite = (0, -6, 0)$ eV, $t = (-5, -3)$ eV and $t_{NNN} = (-3, -2, -2)$ eV.